\documentclass[]{pasj02}

\Received{2025/05/29}
\Accepted{2025/08/27}
\Published{2025/10/07}
 
\usepackage{booktabs}
\usepackage[switch,mathlines]{lineno}
 
\begin{document} 

\title{XRISM High-Resolution X-ray Spectroscopy of Cygnus X-1\\
--- highly ionized Iron absorption structures
}

\author{Shinya Yamada\altaffilmark{1}\thefootnote{*}\orcid{0000-0003-4808-893X}}
\email{syamada@rikkyo.ac.jp}
\author{Natalie Hell\altaffilmark{2,}\thefootnote{*}\orcid{0000-0003-3057-1536}}
\email{hell1@llnl.gov}
\author{Elisa Costantini\altaffilmark{3}\orcid{0000-0001-8470-749X}}

\author{Oluwashina Adegoke\altaffilmark{4}\orcid{0000-0002-5966-4210}}
\author{McKinley Brumback\altaffilmark{5}\orcid{0000-0001-2345-6789}}
\author{Paul Draghis\altaffilmark{5,15}\orcid{0000-0002-2218-2306}}
\author{Ken Ebisawa\altaffilmark{6}\orcid{0000-0002-5352-7178}}
\author{Javier A. Garcia\altaffilmark{4}\orcid{0000-0003-3828-2448}}
\author{Edmund Hodges-Kluck\altaffilmark{7}\orcid{0000-0002-2397-206X}}
\author{Shunji Kitamoto\altaffilmark{1}\orcid{0000-0001-8948-7983}}
\author{Shogo Kobayashi\altaffilmark{8}\orcid{0000-0001-7773-9266}}
\author{Takayoshi Kohmura\altaffilmark{9}\orcid{0000-0003-4403-4512}}
\author{Aya Kubota\altaffilmark{10}\orcid{0000-0002-5413-6304}}
\author{Jon M. Miller\altaffilmark{5}\orcid{0000-0003-2869-7682}}
\author{Misaki Mizumoto\altaffilmark{11}\orcid{0000-0003-2161-0361}}
\author{Tsunefumi Mizuno\altaffilmark{12}\orcid{0000-0001-7263-0296}}
\author{Kaito Ninoyu\altaffilmark{9}\orcid{0009-0003-0640-2828}}
\author{Hiromitsu Takahashi\altaffilmark{13}\orcid{0000-0001-6314-5897}}
\author{Yuusuke Uchida\altaffilmark{9}\orcid{0000-0002-7962-4136}}
\author{Kazutaka Yamaoka\altaffilmark{14}\orcid{0000-0003-3841-0980}}
\author{Sixuan Zhang\altaffilmark{13}\orcid{}}

\altaffiltext{1}{Department of Physics, Rikkyo University, 3-34-1 Nishi Ikebukuro, Toshima-ku, Tokyo 171-8501, Japan}
\altaffiltext{2}{Lawrence Livermore National Laboratory, CA 94550, USA}
\altaffiltext{3}{SRON Netherlands Institute for Space Research, Leiden, The Netherlands} 
\altaffiltext{4}{Cahill Center for Astronomy and Astrophysics, California Institute of Technology, Pasadena, CA 91125, USA}
\altaffiltext{5}{Department of Astronomy, University of Michigan, MI 48109, USA} 
\altaffiltext{6}{Institute of Space and Astronautical Science (ISAS), Japan Aerospace Exploration Agency (JAXA), Kanagawa 252-5210, Japan} 
\altaffiltext{7}{NASA / Goddard Space Flight Center, Greenbelt, MD 20771, USA} 
\altaffiltext{8}{Faculty of Physics, Tokyo University of Science, Tokyo 162-8601, Japan} 
\altaffiltext{9}{Faculty of Science and Technology, Tokyo University of Science, Chiba 278-8510, Japan} 
\altaffiltext{10}{Department of Electronic Information Systems, Shibaura Institute of Technology, Saitama 337-8570, Japan} 
\altaffiltext{11}{Science Research Education Unit, University of Teacher Education Fukuoka, Fukuoka 811-4192, Japan}
\altaffiltext{12}{Hiroshima Astrophysical Science Center, Hiroshima University, Hiroshima 739-8526, Japan} 
\altaffiltext{13}{Department of Physics, Hiroshima University, Hiroshima 739-8526, Japan} 
\altaffiltext{14}{Department of Physics, Nagoya University, Aichi 464-8602, Japan} 
\altaffiltext{15}{MIT Kavli Institute for Astrophysics and Space Research, Massachusetts Institute of Technology, 70 Vassar St, Cambridge, MA 02139, USA}




\maketitle

\begin{abstract}

We present the first high-resolution X-ray spectral analysis of Cygnus X-1 using XRISM.  
The observation was carried out from April 7 to 10, 2024, covering the orbital phase range 0.65--0.17 during its low/hard state.  
Taking advantage of the exceptional energy resolution of the Resolve instrument, 
we examined highly ionized iron absorption lines and characterized the ionization states, column densities, and line-of-sight velocities of the absorbing plasma.  
Spectral analysis revealed an ionization parameter of $\xi \sim 3$, column densities of a few $\times 10^{21}$ cm$^{-2}$, and a blueshifted velocity of $\sim 100$ km~s$^{-1}$.  
The observation was divided into two phases: before and after orbital phase $\phi_{\rm{orb}} = 0.9$, corresponding to non-dipping and dipping intervals.
While only weak absorption features were present before $\phi_{\rm{orb}} = 0.9$, strong absorption by He-like and H-like Fe appeared during the dipping phase.
We measured equivalent widths of 2.3 eV, 0.4 eV, and 1.2 eV for He-like Fe K$\alpha$, and H-like Ly$\alpha_1$ and Ly$\alpha_2$, 
respectively -- demonstrating the capability of XRISM Resolve to securely detect narrow absorption features of only a few eV.
These measurements trace the motion of the absorbing material and offer insight into the kinematics and spatial distribution of the wind in the vicinity of the black hole.  
These findings enhance our understanding of wind-fed accretion in Cygnus X-1 and highlight the importance of continued high-resolution X-ray observations to further constrain the physical properties of winds and accretion flows in high-mass X-ray binaries.

\end{abstract}


\section{Introduction}

X-ray observations provide crucial data on high-energy processes and accretion mechanisms around black holes (BHs). 
Beginning with early X-ray observations of Cygnus X-1 (hereafter Cyg X-1) by the \textit{Uhuru} satellite (e.g., \cite{Oda1971-mb}, \cite{Tananbaum1972-iv}), 
Cyg X-1 is known to be the most massive BH among known high-mass X-ray binaries (HMXBs) in the Milky Way, 
with a BH mass of $21.2 \pm 2.2~M_{\odot}$,
a companion star mass of $40.6^{+7.7}_{-7.1}~M_{\odot}$, 
a distance of 2.2~kpc, and a binary inclination of $i = 27^{\circ}.5^{+0.8}_{-0.6}$ \citep{Miller-Jones2021-nt}.
The companion star is a blue supergiant of type O9.7 Iab, known as HD 226868 \citep{Walborn1973-zj}. 
Matter from the companion star is accreted onto the BH, forming accretion flows. 
Cyg X-1 exhibits two spectral states, the high/soft state and the low/hard state (e.g., \cite{Holt1976-wo}, \cite{Remillard2006-uc}, \cite{Done2007-mj}), 
although it has never shown a purely disk-dominated soft state. 
In the high/soft state, soft X-ray radiation primarily results from multi-blackbody emission of the accretion disk, 
while in the low/hard state, radiation is dominated by inverse Compton scattering in the corona, characterized by a high-temperature electron cloud 
with $kT_{\mathrm{e}} \sim 100$ keV, along with reflection components from the accretion disk (e.g., \cite{Gilfanov1999-jy}, \cite{Makishima2008-wn}, \cite{Tomsick2018-qk}). 
The X-ray timing properties, such as the time-lag of rapid X-ray variability (e.g., \cite{Miyamoto1988-rt}, \cite{Nowak1999-xx}, \cite{Pottschmidt2000-pd}) and frequency space analysis (e.g., \cite{Axelsson2018-df}, \cite{Konig2024-qf}), have also contributed to constraining the physical conditions of the accretion flow.

The HMXB system Cyg X-1/HDE 226868 has gained attention in recent years 
due to the population properties of compact binary mergers inferred from gravitational-wave observations \citep{Abbott2023-yc}. 
Progenitors of binary BH mergers can be BH X-ray binaries with high-mass donor stars. 
Observationally, BH candidates among HMXBs are rare, and those considered to be firmly established are essentially limited to Cyg X-1, LMC X-1, LMC X-3, and M33 X-7.
The formation of such systems, including factors like the Roche lobe filling factor \citep{Hirai2021-ah} 
and the presence of Roche lobe overflow \citep{Xing2025-oh}, 
is the subject of ongoing theoretical studies within the context of binary evolution. 
Nevertheless, there remains a need for observational studies of the fundamental aspects of current binary systems, such as their basic geometry and the mechanisms of mass transfer. 
Recent advancements in X-ray polarization observations, using the Imaging X-ray Polarimetry Explorer (IXPE; \cite{Weisskopf2022-db}), 
have suggested the possibility of an inclination angle of the accretion disk greater than $45^\circ$ \citep{Krawczynski2022-ok}. 
Several alternative ideas have been proposed to reconcile the geometry with a low inclination (e.g., \cite{Dexter2023-op}, \cite{Poutanen2023-vb}, \cite{Steiner2024-hh}). 
Note that since no X-ray eclipse has been observed in this system, the inclination of the orbital plane is considered to be smaller than $67^\circ$ based on the known binary parameters. 
Orbital modulations across optical, UV, X-ray, and radio wavelengths have been detected, caused by changes in the line of sight (LOS). A superorbital period of $\sim$150 days, attributed to the precession or geometry of the disk \citep{Brocksopp1999-hl}, has also been reported, with an alternative interpretation suggesting a period of $\sim$300 days, twice this value \citep{Zdziarski2011-go}.
Additionally, optical polarization observations have suggested the possibility of a tilted accretion disk \citep{Kravtsov2023-ee}. 
Therefore, several aspects of the Cyg X-1 system remain to be further investigated.

\begin{figure*}[h]
  \begin{center}
    \includegraphics[width=\linewidth]{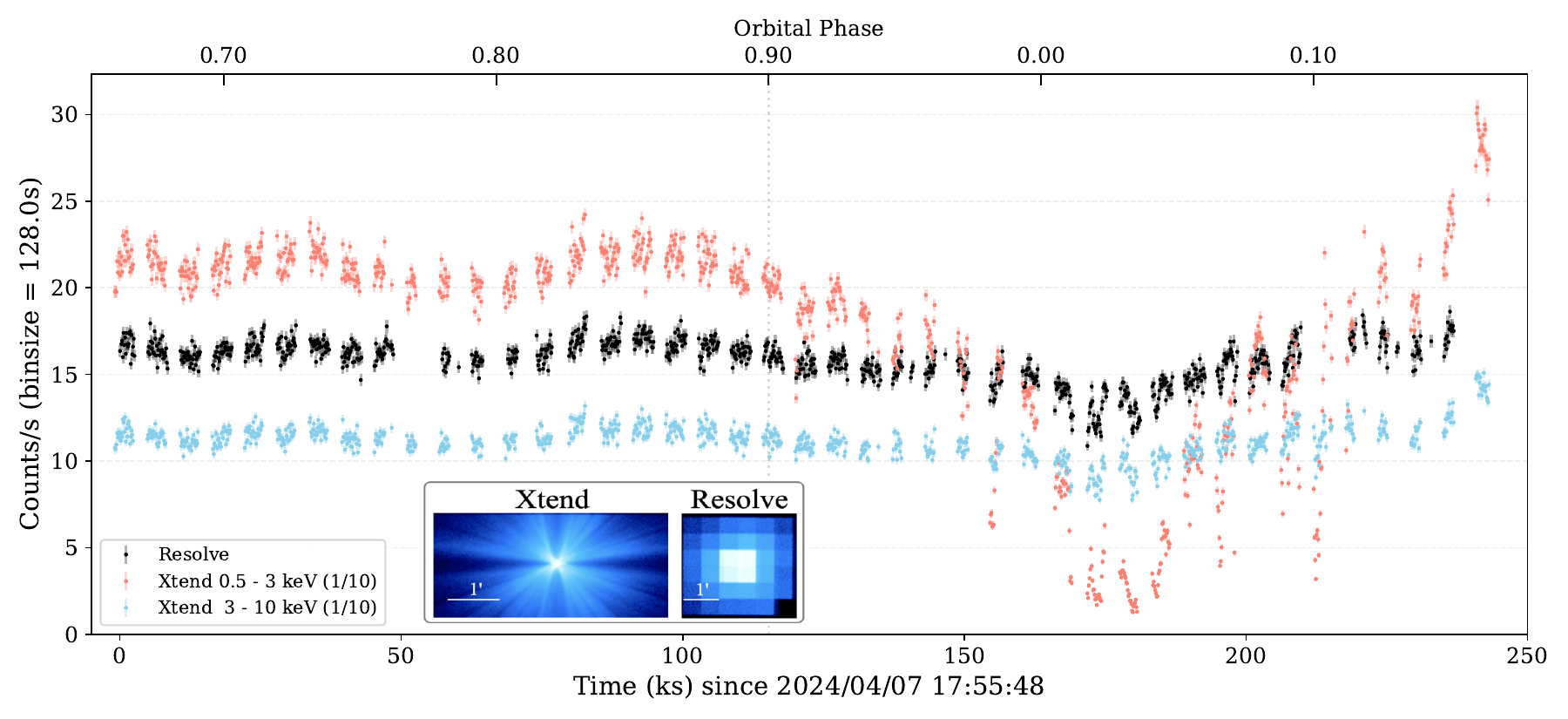}
  \end{center}
  \caption{
Light curves of XRISM and Xtend. The light curve constructed from the entire energy band of Resolve, using only Hp events and all pixels except for the calibration pixel (black), covering approximately 1.6--25 keV. 
The Xtend light curves are obtained in two energy bands: 0.5--3 keV (red) and 3--10 keV (blue).  
The x-axis indicates the elapsed time since the beginning of the observation, and the y-axis shows the count rate with a temporal resolution of 128 s. The Xtend lightcurve is down-scaled by a factor of 10 for visualization purposes. The Xtend and Resolve images from this observation are shown in an inset; note that they have different spatial scales, with fields of view of 3$'\times$3$'$ for Resolve and 38$'\times$38$'$ for Xtend, with a 1$'$ scale bar indicated for reference.
Alt text: The x-axis represents elapsed time since the start of observation, and the y-axis shows count rates with 128-second bins. An inset displays Resolve and Xtend images.}
  \label{fig:lc}
\end{figure*}

Wind accretion in Cyg~X-1 is often described by the Bondi-Hoyle-Lyttleton model \citep{Bondi1944-fo, Bondi1952-hs}, 
though purely quasi-spherical accretion is not fully supported by observations. 
The O-type supergiant companion launches a strong line-driven wind via the CAK mechanism \citep{Castor1975-wm}, 
which is partially accreted onto the BH. 
The large radius of the donor star, likely approaching or marginally underfilling its Roche lobe \citep{Gies1986-iq}, 
suggests that a gas stream may form near the inner Lagrange point (L1). 
The gas flow toward the BH is likely a combination of spherically symmetric wind and Roche lobe overflow, 
often referred to as a focused wind \citep{Friend1982-qb}. 
Recent work by \citet{Ramachandran2025-jn} provides updated stellar parameters based on non-local thermodynamic equilibrium (non-LTE) atmosphere models, 
which may revise the inferred Roche lobe filling factor and mass-transfer geometry in future modeling. 
The dense wind plays a key role in shaping the observed X-ray spectrum and variability. 
Clumping induced by radiative instabilities, the Coriolis effect, and X-ray irradiation from the compact object 
can significantly influence wind structure and dynamics, as demonstrated by radiation hydrodynamic simulations (e.g., \cite{Cechura2015-cc}, \cite{Sundqvist2018-fh}). 
Episodes of enhanced absorption, or ``dips'' -- short-term decreases in X-ray flux -- occur near the superior conjunction of the BH, 
when the observer, donor star, and BH are aligned (e.g., \cite{Pravdo1980-il}, \cite{Kitamoto1984-hc}), 
corresponding to an orbital phase of $\phi_{\rm{orb}} = 0$. 
In this paper, we adopt an orbital period of 5.599829 days and an ephemeris of 41874.207 $\pm$ 0.009 MJD \citep{Brocksopp1999-hl}.

Long-term monitoring with \textit{RXTE} has shown that absorption dips are predominantly observed in the low/hard state, while they are rare or absent in the high/soft state \citep{Wen1999-iw}. Comprehensive studies using \textit{RXTE} data \citep{Balucinska-Church2000-kb, Boroson2010-oo} have mapped the orbital distribution of dips, showing that they occur most frequently around superior conjunction ($\phi_{\rm orb} \sim 0$). 
Notably, \citet{Balucinska-Church2000-kb} reported that the distribution is asymmetric with respect to phase zero, peaking around $\phi \sim 0.95$. 
\citet{Grinberg2015-kl} extended these studies by examining long-term variability and dip behavior across different spectral states.
Detailed studies of absorption features have been performed using \textit{Chandra} HETG observations of Cyg~X-1 at various orbital phases and spectral states: 
e.g., $\phi_{\rm{orb}} \approx 0.74$ \citep{Schulz2002-jp} and $\phi_{\rm{orb}} \approx 0$ \citep{Hanke2009-rh} in the low/hard state, 
$\phi_{\rm{orb}} \approx 0.77$ \citep{Miller2005-sw} in the intermediate state, 
and $\phi_{\rm{orb}} \approx 0.88$ \citep{Chang2007-qp} and $\phi_{\rm{orb}} \approx 0$ \citep{Feng2003-hu} under conditions similar to the high/soft state.
\citet{Miskovicova2016-zi} conducted a detailed analysis of five non-dip spectra during the low/hard state at various orbital phases 
($\phi_{\rm{orb}} \approx 0.0$, 0.2, 0.5, 0.75), detecting numerous highly ionized absorption lines. 
They reported that the line profiles vary systematically with orbital phase: pure, symmetric absorption lines appear near superior conjunction ($\phi_{\rm{orb}} \approx 0$), 
while P Cygni profiles become prominent around inferior conjunction ($\phi_{\rm{orb}} \approx 0.5$), indicating wind asymmetries and ionization structure.
Complementary to this, \citet{Hirsch2019-em} analyzed dip spectra obtained during the low/hard state to characterize the physical properties of the denser wind clumps responsible for the absorption dips. By tracking changes in the Si and S line regions (1.6--2.7 keV), they found lower ionization stages in deeper dip phases, suggesting that the clumps show variations in ionization state due to density inhomogeneities.
Using \textit{XMM-Newton} observations during the low/hard state, and applying the clumpy wind model proposed by \citet{El-Mellah2020-sy}, 
the stellar mass-loss rate has been estimated to be $7 \times 10^{-6}~M_{\odot}~\mathrm{yr^{-1}}$ \citep{Lai2024-fu}. 
However, the detailed properties of the absorbing material remain poorly understood, 
partly due to the limited effective area of previous instruments in the Fe-K band at high spectral resolution.

In this paper, we report on the first high-resolution spectroscopy of Cyg~X-1 
with high photon statistics in energies above 2~keV, enabled 
by the observation of the X-ray Imaging and Spectroscopy Mission (XRISM; \cite{Tashiro2020-iz}).
XRISM, launched on 2023 September 7, carries two co-aligned instruments, Resolve (\cite{Ishisaki2022-ci}, \cite{Sato2023-gr}, \cite{Scott2025}) 
and Xtend (\cite{Mori2022-wo}, \cite{Noda2025-aq}, \cite{Uchida2025-ld}), 
located at the focal plane of two X-ray Mirror Assemblies (XMAs; \cite{Tamura2024-wb}). 
This paper focuses on the velocity structure and thermal properties of the hot plasma that can be constrained by analyzing the well-resolved, 
absorption lines detected in the latter half of the observation. 
The results provide insights into the nature of the stellar wind and its environment, 
which help to investigate the evolution of the Cyg X-1 system. 
Other investigations, including a search for weak emission features and broadband spectroscopy, are left for future work. 
This paper is organized as follows. In section 2, we describe the details of the observations and data reduction. 
In section 3, we present the detailed spectral analysis. The implication of the results are discussed in section 4. 
Finally, we conclude this study in section 5. 
Given that this is one of the first papers on the scientific outcome of Cyg X-1 with XRISM, 
we provide a brief descriptions about the gain calibration and the need for special treatment for the bright source in the appendices. 
BH spin is a key observational quantity, as investigated in various previous studies (e.g., \cite{Draghis2024-oz}, \cite{Zdziarski2024-mi}). 
Due to the need for careful treatment of the continuum component, its analysis is deferred to a future paper.
The errors quoted in the text and tables, and the error bars given in the figures represent the 1$\sigma$ confidence level, unless otherwise stated.

Note that \citet{Ramachandran2025-jn} reported lower values for the BH mass ($\sim$13--18 $M_{\odot}$) and the donor star mass ($\sim$29 $M_{\odot}$), 
compared to those in \citet{Miller-Jones2021-nt}. 
These revised estimates are based on the \texttt{PoWR} stellar atmosphere models (e.g., \cite{Grafener2002-ib}), 
which solve the radiative transfer and hydrodynamic equations under non-LTE conditions and account for expanding, line-driven stellar winds in a self-consistent manner. 
Such models offer a more realistic treatment of ionization balance and wind acceleration than conventional LTE, plane-parallel models. 
While these advancements warrant careful consideration in future studies, 
we adopt the parameters from \citet{Miller-Jones2021-nt}, including LTE models in \citet{Orosz2011-ez}, in this work to maintain consistency with previous observational analyses and model comparisons.

\begin{figure*}[h]
  \begin{center}
    \includegraphics[width=\linewidth]{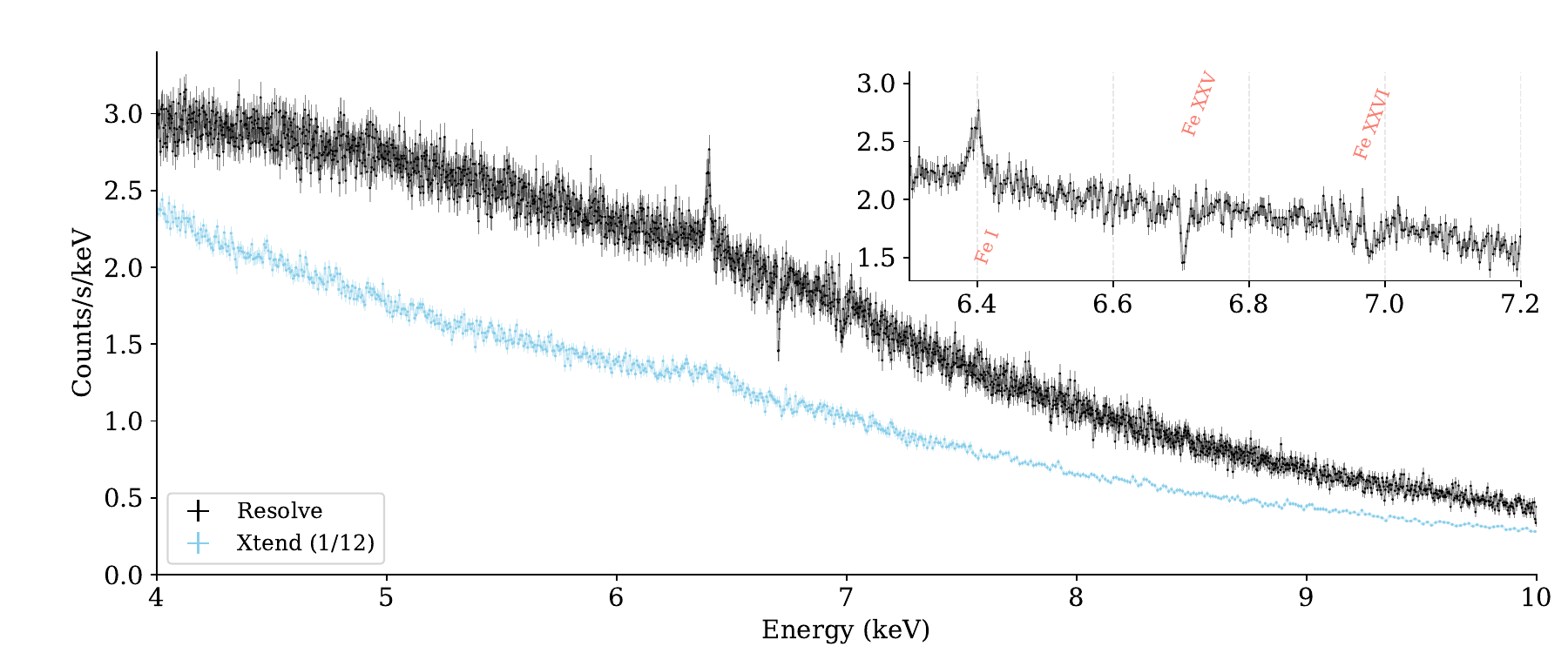} 
  \end{center}
  \caption{Observed spectra of Cyg X-1 in the 4--10 keV range from Resolve and Xtend. Both spectra include the effects of detector response.  
The shape of the Resolve spectrum is primarily influenced by absorption through the beryllium window in the gate valve.  
The vertical axis of the Xtend spectrum is scaled down by a factor of 12.  
The inset in the upper right corner provides a close-up view of the Resolve spectrum in the 6.3--7.2 keV range,  
highlighting fluorescence emission from neutral Fe at 6.4 keV and absorption lines from $n=1 \rightarrow 2$ transitions in He-like and H-like Fe at $\sim$6.7 keV and $\sim$7.0 keV, respectively.  
Alt text: Observed spectra of Cyg X-1 from XRISM Resolve and Xtend in the 4--10 keV band.  
Inset shows a close-up of the 6.3--7.2 keV range in the Resolve spectrum.}
  \label{fig_spec}
\end{figure*}

\section{Observation and Data Reduction}

\subsection{General properties of the observation}

The XRISM observation of Cyg~X-1 was conducted from 2024-04-07 16:55:04 to 2024-04-10 13:41:04 (UTC).
The orbital phase spans from 0.654 to 0.167 when the phase is calculated in MJD (cf.~0.671 to 0.183 in HJD).
Out of the total observation duration of $\sim$248~ks, $\sim$125~ks corresponds to the effective exposure time,
primarily affected by Earth occultations and passages through the South Atlantic Anomaly (SAA).
The observation ID (OBSID) is 300049010. 
Data reduction was performed using the JAXA pre-pipeline software version ``004 002.15Oct2023 Build7.011'', 
the pipeline script ``03.00.013.009'', and the CALDB8 database, with files ``gen20240315\_xtd20240815\_rsl20240815''.

During the 2024 April 7--10 observation window, the MAXI monitor \citep{Mihara2011-dd, Matsuoka2009-as} reported 
2--4~keV, 4--10~keV, and 10--20~keV fluxes of $\sim$0.50, 0.40, and 0.14 photons~s$^{-1}$~cm$^{-2}$, respectively 
(values retrieved from the public archive on April 8, 2024).
The corresponding hardness ratio (4--10~keV / 2--4~keV) was $\sim$0.8, which exceeds the threshold of 0.48 for the hard state defined by \citet{Sugimoto2016-pc}.
In addition, according to the state-classification scheme of \citet{Grinberg2013-jd}, 
the Swift/BAT \citep{Krimm2013-or} 15--50~keV count rate was $\sim$0.16~counts~s$^{-1}$~cm$^{-2}$, 
a level characteristic of the hard (or possibly intermediate) state.
Taken together, these monitoring data confirm that Cyg~X-1 was in the low/hard state during the XRISM observation.
 
\subsection{Data Reduction of Resolve}

The Resolve observation of Cyg X-1 was conducted through a $\sim$250~$\mu$m-thick beryllium window \citep{Midooka2020-ns} in the closed aperture door, which limited the observable energy range to above 1.6~keV. To suppress pile-up and grade migration caused by the brightness of the source, the neutral density (ND) filter was employed. This filter consists of a 0.25~mm-thick molybdenum plate with 1.1~mm-diameter perforations, resulting in a total throughput of 24.5\% \citep{de-Vries2017-oo, Shipman2024-tr}. The molybdenum fluorescence lines at 17.48~keV (Mo K$_{\alpha 1}$) and 2.3~keV (Mo L$_{\alpha 1}$) did not appear in the spectrum.
The Resolve data were screened using \texttt{xa\_gen\_select\_20190101v004.fits} to remove time intervals affected by the Earth’s eclipse, the bright limb, passages through the South Atlantic Anomaly (SAA), and recharging cycles of the 50~mK cooler operated by the ADR (adiabatic demagnetization refrigerator) system \citep{Shirron2018-wf}.

Our analysis involved examining standard clean events that were processed through the instrumental pipeline, starting from raw data and subsequently handled by the pipeline software (version 1.0). Source events were extracted from all pixels except the calibration pixel (pixel 12). Although four central pixels (pixel 0, 17, 18, 35) were especially bright, no event loss due to the PSP limit occurred, and all events were recorded in orbit. The background level across all pixels was negligible given the brightness of Cyg X-1. We applied standard screening based on pixel coincidence and an energy-dependent rise time cut. The calibration pixel was used solely for energy scale calibration. After screening, the net exposure time was 125.6~ks.

The brightness and short-term variability of Cyg X-1 in the low/hard state necessitate a careful understanding of how events are recorded and graded in the Resolve detector system. The signal processing chain -- largely inherited from the ASTRO-H SXS design \citep{Kelley2016-xk} -- includes a detector array, Xbox electronics (\cite{Porter2010-ae}; \cite{Porter2018-ip}), and a Pulse Shape Processor (PSP) \citep{Ishisaki2018-wt}. Absorbed X-ray photons generate thermal pulses in the HgTe absorbers, which are amplified, digitized, and analyzed using an optimal filtering technique (``matched filter''; \cite{Szymkowiak1993-md}). The PSP evaluates pulse heights and assigns event grades based on the temporal proximity of neighboring pulses \citep{Tsujimoto2017-rc}. Events are categorized as ``high primary'' (Hp) when temporally isolated within 1024 samples (81.92~ms), or as ``medium primary'' (Mp) when accompanied by other pulses within shorter separations.
At high count rates, overlapping untriggered pulses can degrade energy resolution. The ND filter mitigates this by limiting total throughput. 
In addition, as illustrated in Appendix \ref{sec:grade}, the count rate of Hp events in some pixels does not scale linearly with the instantaneous brightness of Cyg X-1, 
due to an increased fraction of secondary events.
While the inclusion of Mp events can increase the photon statistics, a spectral analysis including Mp events will be addressed in future work.

\begin{figure*}[h]
  \begin{center}
    \includegraphics[width=\linewidth]{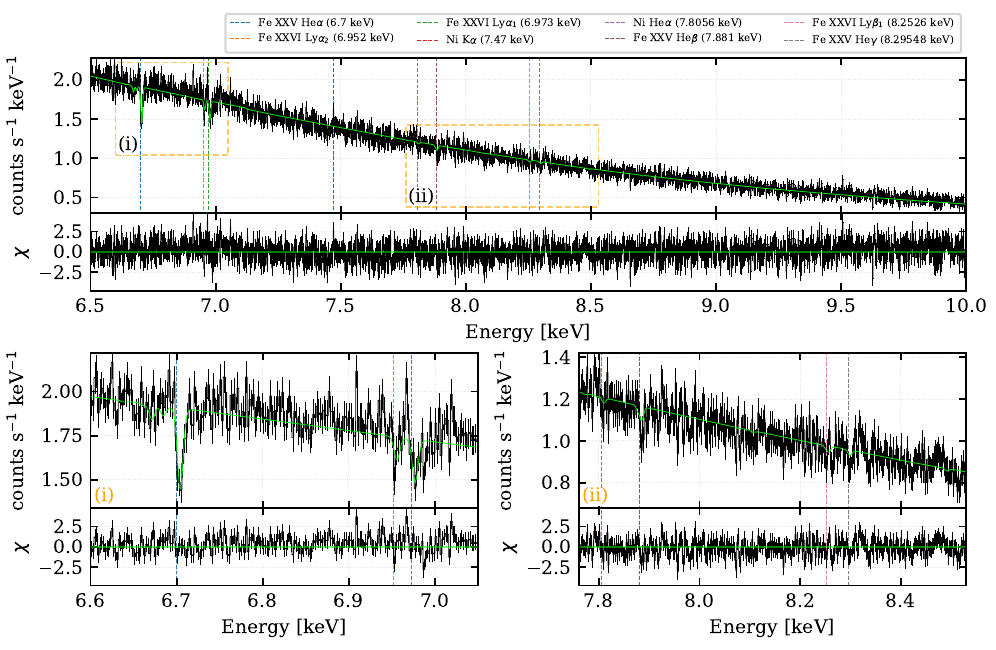} 
  \end{center}
  \caption{Time-averaged Resolve spectrum and the best-fit model, with enlarged views highlighting the He-like and H-like Fe K$\alpha$ and K$\beta$ series.  (Top) Measured photon count spectrum without any correction for effective area and the best-fit model over 6.5--10~keV. (Bottom) A close-up view at around Fe K${\alpha}$ (left) and Fe K${\beta}$ (right). For each panel, residuals between the data and the model are shown. 
  The energies of representative absorption lines are indicated by vertical auxiliary lines, with their meanings described in the legend.
Alt text: Energy v.s., time-averaged XRISM Resolve spectrum of Cyg X-1. Residuals between the data and the model are shown for each panel.}
\label{fig_avespec_fit}
\end{figure*}

During this observation, the average Hp count rate was 15 c s$^{-1}$ per array.  
With the ND filter reducing the incident X-ray flux to one-fourth, 
Hp events remain the dominant events even at the central four pixels.
Spectral and imaging analysis with Resolve were conducted with several preprocessing steps, 
including generating the Redistribution Matrix File (RMF), exposure map, and Ancillary Response Files (ARF).
The RMF, which defines energy-to-channel conversion, energy resolution, branching ratios, and detector response \citep{Eckart2018-bc}, 
was created using \texttt{rslmkrmf} with the ``Large'' option, the cleaned event file, and parameter file \texttt{xa\_rsl\_rmfparam\_20190101v006.fits}. 
Line-spread function components included the Gaussian core \citep{Mourice2025}, exponential low-energy tail, escape peaks, and silicon K$\alpha$ 
fluorescence lines. The exposure map, accounting for pixel-specific effective exposure time, was generated using \texttt{xaexpmap}.
The ARF, an effective area of the X-ray mirror with filter effects, was produced using \texttt{xaarfgen} for specified regions. 
This included the GV closed configuration, assuming a point source at the aim point, 
and the calibration database of XMAs (\cite{Tamura2022-sf}, \cite{Boissay-Malaquin2022-rz}, \cite{Hayashi2022-tw}, and \cite{Boissay-Malaquin2024-bz}).   

\subsection{Data Reduction of Xtend}

During this observation, Xtend operated in \texttt{WINDOW2BURST} mode, in which the CCD is read out for 0.0620352~s every 0.5~s (1/8 window mode). After standard event screening, the effective exposure time was 15.09~ks, corresponding to the burst-to-frame time ratio applied to the full observation duration ($\simeq$125~ks~$\times$~0.062/0.5). 
The Xtend instrument, consisting of a 2$\times$2 array of P-channel back-illuminated CCDs, is paired with the X-ray Mirror Assembly (XMA) to provide a wide field of view of 38$'$$\times$38$'$. The CCDs incorporate a 200~$\mu$m-thick depletion layer, which enhances quantum efficiency (QE) and suppresses non-X-ray background (NXB) around 6~keV by shifting the Bragg peak energy of cosmic rays beyond the instrumental band. Compared to the 40~$\mu$m-thick BI CCDs onboard Suzaku/XIS \citep{Koyama2007-jq}, Xtend achieves significantly lower NXB levels. Optical leakage is mitigated by an optical blocking layer (OBL) composed of an $\sim$200~nm-thick aluminum film, which helps reduce pseudo events. The operational bandpass, extending from 0.4 to 13.0~keV, has been empirically confirmed \citep{Noda2025-aq}.

For this bright source, careful treatment of pile-up effects was necessary. Based on the empirical criteria by \citet{Yamada2012-qx}, with modifications for Xtend, we estimate a pile-up fraction of $\sim$3\% within a 20$''$ radius ($\sim$10 pixels). While this effect may distort the continuum, narrow line features are relatively unaffected. Therefore, we extracted spectra using a circular region of 60 pixels (1 pixel = 1.78$''$) to retain high photon statistics. 
Although flickering pixels -- defined as pixels that occasionally produce events unrelated to genuine X-ray or particle interactions -- can appear in CCD data, the \texttt{xtdflagpix} tool was not applied in this analysis. This decision was based on the fact that, for bright sources like Cyg X-1, flickering pixel events are typically minor compared to the dominant astrophysical signal. A similar approach was adopted for bright-source observations with Suzaku/XIS. The validity of this choice was confirmed by comparing spectra extracted from different regions, where no statistically significant differences beyond the expected uncertainties were observed.

The Xtend image is shown in an inset in Figure~\ref{fig:lc}.  
The RMF was generated using \texttt{xtdrmf} and includes line-spread function components such as the Gaussian core, exponential tail, silicon escape peaks, and Si K$\alpha$ fluorescence \citep{Inoue2016-kk}.  
The exposure map was produced using \texttt{xaexpmap}, and the ARF was computed with \texttt{xaarfgen} assuming a point source within a 60-pixel circular region.  
We did not subtract the NXB in this analysis, as it remained below 1\% of the source signal across the entire energy range.  
The Xtend data were used to estimate the overall source flux and served as input for the photoionization model calculations.

\subsection{Lightcurves}

The light curves of Resolve and Xtend were created with a time bin size of 128~s, as shown in Figure~\ref{fig:lc}. 
For Resolve, the light curve represents the count rate of Hp events without accounting for branching ratios. 
The light curve was corrected for effective exposure in each time bin, accounting for lost time due to Good Time Interval (GTI) gaps and readout time for Xtend. 
The vertical axis of the Xtend light curve is scaled down by a factor of 10, primarily due to the Gate Valve (GV) closed mode and the use of a Neutral Density (ND) filter applied to Resolve.
In the 0.5--3~keV band, the Xtend count rate remained approximately constant at $\sim 200~\mathrm{c~s^{-1}}$ during the early part of the observation, 
followed by a sharp dip to $\sim 20~\mathrm{c~s^{-1}}$ around $\phi_{\mathrm{orb}} \simeq 0.03$. 
Subsequently, the count rate gradually increased again, reaching $\sim 140~\mathrm{c~s^{-1}}$ by $\phi_{\mathrm{orb}} \sim 0.15$. 
This recovery likely indicates that the source emerged from deep absorption, possibly combined with intrinsic spectral changes.

The 3--10~keV band showed relatively moderate variations: it began at $\sim110~\mathrm{c~s^{-1}}$, declined to $\sim90~\mathrm{c~s^{-1}}$ around $\phi_{\mathrm{orb}} \sim0.05$, 
and then gradually increased to $\sim130~\mathrm{c~s^{-1}}$ toward the end of the observation. 
These variations are consistent with the trend observed in Resolve. 
The pronounced dip behavior in the soft X-ray band around $\phi_{\mathrm{orb}} \approx 0$ is attributed to spatially localized, less-ionized absorbing material—commonly referred to as ``dips''. 
In contrast, the more gradual spectral changes observed over the broader timescale likely reflect intrinsic variability of the central source. 
To investigate the spectral evolution as a function of orbital phase, we defined the non-dip phase as $\phi_{\mathrm{orb}} < 0.9$ 
and the dipping phase as $\phi_{\mathrm{orb}} \geq 0.9$, and analyzed the corresponding spectra in detail.

\subsection{Wide-band spectra}

To provide an overview of the XRISM data for Cyg X-1 obtained in this observation, Figure~\ref{fig_spec} shows the broadband spectra spanning 4--10~keV. 
We note that the 1.6--4~keV band, although available, is not included in this figure, as a detailed analysis in that range will be presented in a forthcoming paper.
The Resolve spectrum represents the measured photon count spectrum without any correction for effective area.
The Xtend spectrum is shown over the same energy range, with its vertical axis scaled down by a factor of 1/12 for visibility.
The apparent differences in spectral shape between Resolve and Xtend primarily reflect the distinct detector architectures and materials. Resolve employs cryogenic microcalorimeters with HgTe absorbers, while Xtend uses conventional Si-based CCDs. These intrinsic differences in energy-dependent absorption efficiency and spectral redistribution affect the observed count spectra even for the same source flux. For Resolve in particular, additional components along the optical path influence the detection efficiency. 
These include five thin-film optical blocking filters (Al/polyimide), the beryllium window of the gate valve, and the ND filter used during this observation.
These materials are incorporated into the response matrix used in the analysis \citep{Eckart2018-bc}.

The superior energy resolution of Resolve is evident from the sharp fluorescence X-ray line observed at $\sim$6.4~keV, attributed to neutral Fe (Fe~I). 
Although the spectral resolution is in principle sufficient to resolve the K$\alpha_1$ and K$\alpha_2$ components, as demonstrated using the Fe-55 calibration source in Appendix~\ref{sec_fe55}, the observed line profile shows only a shallow dip between the two peaks. 
This may suggest broadening due to orbital Doppler shifts or turbulence within the system, but a detailed analysis of the line profile and its time variability will be presented in a subsequent paper.  While there is a marginal indication of Ni K$\alpha$ emission at $\sim$7.48~keV, its statistical significance is not sufficient to claim a robust detection. 
The Xtend spectrum, with an energy resolution of $\sim$180~eV \citep{Mori2024-nu}, shows a 6.4~keV spectral structure consistent with previous observations 
using Suzaku/XIS (e.g., \cite{Makishima2008-wn}). 
In the Resolve spectrum, absorption lines corresponding to the $n=1 \rightarrow 2$ transitions in He-like and H-like Fe are also detected. 
These features have equivalent widths of only a few eV, making them difficult to detect in the Xtend data. 
Since the observation includes orbital phases near $\phi_{\rm{orb}} \sim 0$, when the stellar wind is expected to be denser, 
the origin of these absorption lines is likely related to the wind geometry. 
The ability to detect such weak absorption features of highly ionized Fe highlights the advantages of Resolve's high energy resolution and sensitivity.

\begin{figure}[h]
  \begin{center}
    \includegraphics[width=\linewidth]{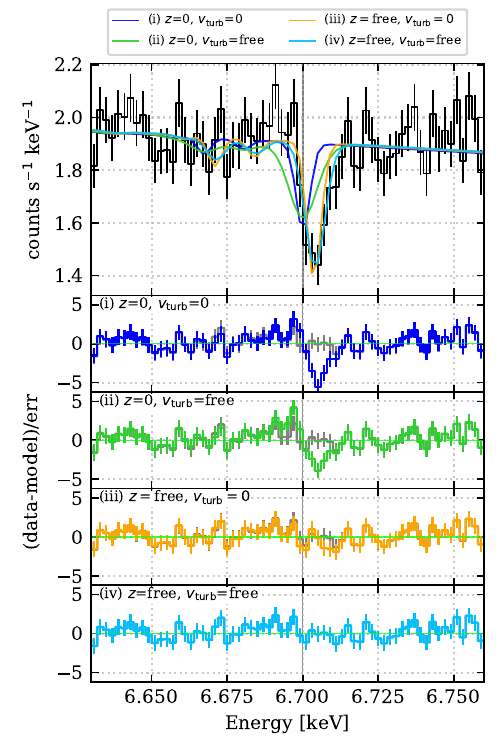} 
  \end{center}
  \caption{Detailed view of the measurement of turbulent velocity and Doppler shift using the time-averaged spectrum fitted with an ionized absorption model.  
(Top) Comparison of four representative models applied to the time-averaged spectrum. The solid lines correspond to:  
(i) $z = 0$, $v_{\text{turb}} = 0$ (blue),  
(ii) $z = 0$, $v_{\text{turb}}$ free (green-yellow),  
(iii)$z$ free, $v_{\text{turb}} = 0$ (yellow), and  
(iv) $z$ free, $v_{\text{turb}}$ free (cyan).  
(Bottom) Panels (i)--(iv) show the residuals between the data and each best-fit model. For comparison, the residuals of model (iv) are also overplotted in gray.  
Alt text: Energy versus the spectrum around 6.7 keV, and residuals from various models.
}
\label{fig_avespec_fit_fine}
\end{figure}

\begin{table*}[h]
\caption{Fitting results of the single-zone absorber model for the time-averaged spectra over 6.5 -- 10 keV, including four cases: turbulent velocity and redshift parameters fixed at 0, and both allowed to vary freely.}
  \centering
  \begin{tabular}{ccccc}
  \hline\hline
      paramater & $v_{\text{turb}}$=0, z=0 & z=0 & $v_{\text{turb}}$=0 & $v_{\text{turb}}$, z= free  \\ \hline
      $\xi$ & 3.21$\pm$ 0.06 & 3.16$_{-0.06}^{+0.05}$ & 3.20$\pm$ 0.03 & 3.19$\pm$ 0.03  \\ 
      $N_{\rm H} $$^a$ [$10^{22}$ cm$^{-2}$] & 0.19$_{-0.04}^{+0.05}$ & 0.26$_{-0.06}^{+0.05}$ & 0.32$\pm$ 0.04 & 0.34$_{-0.04}^{+0.06}$  \\[0.3em]
      $z \times 10^{-4}$ & 0 (fix) & 0 (fix) & $-5.5\pm 0.5$ & $-6.0\pm 0.6$  \\ 
      $cz$$^b$ [km s$^{-1}$] & 0 (fix) & 0 (fix) & $-164.9\pm 15.0$ & $-176.9\pm 18.0$  \\[0.3em]  
      $v_{\rm{turb}}$ [km] & 0 (fix) & 260$_{-56}^{+60}$ & 0 (fix) & 139$_{-29}^{+37}$  \\
      Photon Index & 1.975$\pm$ 0.013 & 1.982$\pm$ 0.013 & 1.982$_{-0.012}^{+0.013}$ & 1.986$\pm0.013$  \\ 
      norm$^c$ & 1.93$\pm$ 0.05 & 1.96$\pm$ 0.05 & 1.96$\pm$ 0.05 & 1.97$\pm0.05$  \\ 
      \hline
      $\chi^2$/dof & 1984.6/1745 & 1968.6/1744 & 1895.7/1744 & 1884.8/1743 \\
  \hline\hline
  \end{tabular}
  \label{table_avespec_fit}    

\begin{small}
\begin{itemize}
\setlength{\parskip}{0cm} %
\setlength{\itemsep}{0cm} %
\item[$^a$] This is a column density for the ionized absorber. Note that the neutral column density is fixed at 6.0$\times$10$^{21}$ cm$^{-2}$. 
\item[$^b$] The free parameter is $z$, while $cz$ is presented to help understand the velocity unit. 
\item[$^c$] The normalization of the powerlaw model in a unit of photons keV$^{-1}$ cm$^{-2}$ s$^{-1}$ at 1 keV. 
\end{itemize}
\end{small}
\end{table*}

\section{Data Analysis and Results}

\subsection{Time-averaged spectra}
\label{sec:timeave}

\begin{figure*}[h]
  \begin{center}
    \includegraphics[width=0.99\linewidth]{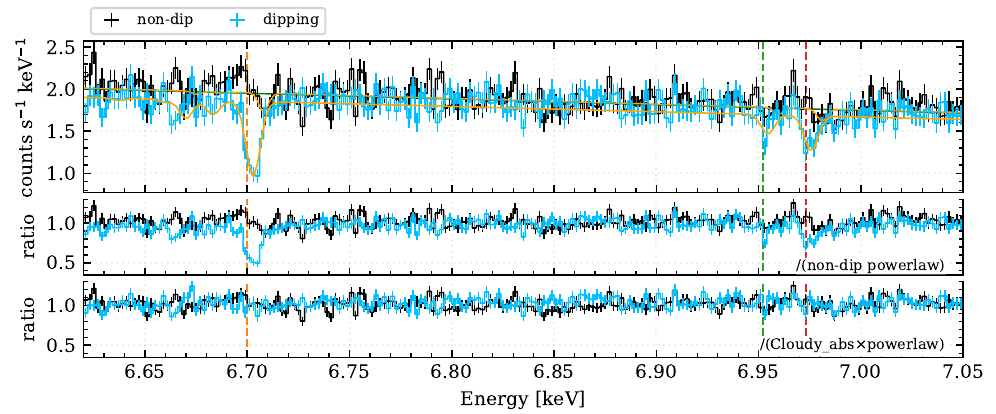} 
  \end{center}
  \caption{
 (Top) 
 The measured photon count spectra divided at $\phi_{\rm{orb}}=0.9$. The data during dipping intervals are shown in light blue, while the non-dip spectrum are shown in black. The best-fit models (solid yellow lines) and the power-law continuum (solid green line) fitted to the 4--10 keV spectrum of Cyg X-1 are overlaid.   
(Middle) Ratios of the spectra to the powerlaw model (green solid in the top panel). 
(Bottom) Ratios of the spectra to the best-fit ionized absorption models for both non-dip and dipping intervals. 
The parameters are presented in Table~\ref{table_dip_nondip}.
The vertical auxiliary lines have the same meaning as in Figure~\ref{fig_avespec_fit}. 
Alt text: Energy v.s. the spectra of Cyg X-1 divided at orbital phase.}
\label{fig_dip_nondip}
\end{figure*}

In this paper, we aim to report the observational results of absorption lines from highly ionized Fe. Detailed analyses of time variability, line profiles, and other specific characteristics will be addressed in subsequent publications. As a primary focus, we limit our discussion to what can be explained under the assumption of a single absorption zone and a single continuum component. Consequently, the energy range analyzed is restricted to 6.5--10 keV. For the continuum, a single power-law model is adopted. 

Regarding the absorption model, we utilized multiple approaches, including widely used ionized absorption models such as \texttt{Cloudy} (\cite{Chatzikos2023-ym}, \cite{Gunasekera2023-vr}), \texttt{xstar} \citep{Kallman2001-au}, and the \texttt{pion} model from SPEX \citep{Mehdipour2016-cg}, as well as a phenomenological model, \texttt{ion\_abs} (\cite{Ueda2004-tv}, \cite{Tomaru2020-cw}), which was generated using Voigt profiles, a simplified elemental composition, and an absorption line database. Since these models showed no significant differences in the results, this paper primarily presents the results based on the \texttt{Cloudy} model. Minor differences among the models will be discussed in future papers.

We first present the results of the analysis using the entire observation period, during which the statistics are maximized. 
The absorption model employed is a table model for XSPEC generated by \texttt{Cloudy}. 
The spectral energy density (SED) used as input for \texttt{Cloudy} was constructed based on the broadband X-ray continuum 
(in units of photons cm$^{-2}$ s$^{-1}$ keV$^{-1}$) measured during the XRISM observation window, incorporating data from Xtend as well as a simultaneous NuSTAR observation 
(ObsID: 30901039002, conducted from 2024-04-08 18:16 to 2024-04-09 06:02 UTC). 
The NuSTAR data were used to constrain the high-energy portion of the continuum, particularly the spectral slope and cutoff energy.

The continuum is modeled in XSPEC using the \texttt{diskbb} + \texttt{nthComp} components, with the following parameters:  
for \texttt{diskbb}, $T_{\rm in} = 0.166~\mathrm{keV}$ and $\mathrm{norm} = 6.11 \times 10^5$;  
for \texttt{nthComp}, $\Gamma = 1.67$, $kT_e = 100~\mathrm{keV}$, $kT_{\rm bb} = 0.166~\mathrm{keV}$, 
$\mathrm{inp\_type} = 0$, $\mathrm{redshift} = 0$, and $\mathrm{norm} = 1.50$.  
This model yields an unabsorbed flux in the 0.1--500~keV range of $4.9 \times 10^{-8}$~erg~cm$^{-2}$~s$^{-1}$,  
corresponding to a luminosity of $L \sim 3 \times 10^{37}$~erg~s$^{-1}$ and an Eddington ratio of $L/L_{\text{Edd}} \sim 1\%$, 
which is typical for the low/hard state of Cyg~X-1.  
We designate the absorption model generated with this SED as \texttt{cloudy\_abs}.
Details of the broadband spectral modeling, including the full NuSTAR data, will be reported in a separate publication.

Using this model, the data were fitted with a conventional model implemented in \texttt{XSPEC}, 
i.e., the \texttt{tbabs} * \texttt{cloudy\_abs} * \texttt{powerlaw} configuration to quantify the absorption features. 
The interstellar absorption was modeled using \texttt{tbabs} with abundances from \citet{Wilms2000-nv}, 
and the hydrogen column density was fixed at $6.0 \times 10^{21}$~cm$^{-2}$. 
This value is slightly higher than the line-of-sight interstellar column toward Cyg~X-1 
measured using X-ray dust scattering halos by \citet{Xiang2011-da}, who reported $N_{\rm H} \sim 4 \times 10^{21}$~cm$^{-2}$, 
but is consistent with values derived from previous X-ray spectral studies such as \citet{Makishima2008-wn}, who found $6.6 \times 10^{21}$~cm$^{-2}$ using Suzaku. 
We note that our analysis focuses on the 6.5--10~keV band, where the effect of interstellar absorption does not significantly impact the results on the absorption features.


The \texttt{cloudy\_abs} model has four parameters: the column density of the ionized absorber ($N_{\rm H}$), the turbulent velocity ($v_{\rm turb}$), the redshift ($z$), and the ionization parameter ($\xi$), which is defined as $\xi = L / (n r^2)$ \citep{Tarter1969-jj}, where $L$, $n$, and $r$ denote the X-ray luminosity, particle density, and distance from the source, respectively.
In \texttt{Cloudy}, the ionization parameter is calculated using the integrated luminosity over the energy range of 13.6 eV to 13.6 keV. 
The $\xi$, iron ion distributions, and electron temperatures calculated using this SED are presented in the Appendix \ref{sec_cloudy}. 
When $\log (\xi)$ reaches $\sim$3, a significant population of H-like and He-like Fe ions is present, and the electron temperature reaches $\sim 10^7$ K.

When the emission from the O star is included in the SED for photoionization -- 
modeled using \texttt{bbody} with $kT = 2.4 \times 10^{-3}$ keV and $\text{norm} = 47.52$ -- 
the value of $L$ is inflated due to the contribution of UV photons.  
Although these photons do not significantly contribute to the production of highly ionized iron, they are still included in the calculation of $L$.  
This leads to a higher value of $\xi$, even though these UV photons cannot ionize electrons in the Fe K-shell.  
Since we confirmed that the results presented in this paper are not affected by the inclusion of O-star emission,  
the input SED for the photoionization calculation does not include it.  
Because the velocity involved is much less than the speed of light $c$,  
the redshift $z$ is approximated as $v/c$, where $v$ is the line-of-sight (LOS) velocity, and a negative $v$ indicates motion toward the observer.

We begin by presenting the absorption lines in detail.  
Figure~\ref{fig_avespec_fit} shows the spectral fit in the 6.5--10~keV range. 
Panel (i) displays a close-up view of the He-like Fe K$_{\alpha}$ ($\sim$6.7~keV) and H-like Fe K$_{\alpha}$ ($\sim$7.0~keV) lines, both of which are statistically significant even when fitting a narrower energy range. 
The He-like Fe K$_\beta$ absorption line is also clearly detected, while the H-like Fe K$_\beta$ line is only marginally visible in panel (ii).  
The best-fit parameters are summarized in Table~\ref{table_avespec_fit}. 
The results shown in Figure~\ref{fig_avespec_fit} correspond to a fit in which both the turbulent velocity and redshift were allowed to vary.  
The ionization parameter was found to be $\log \xi \sim 3.2$, with a column density of $3 \times 10^{21}$~cm$^{-2}$.  
The derived photon index is not physically meaningful in itself, as the fit simply models the narrow energy band with a power-law function.  

We note that the Fe XXVI Ly$\alpha$ absorption profile near 6.95--7.00~keV appears slightly asymmetric and more complex than predicted by a simple absorption component. 
This suggests that the observed line shape may be influenced by multiple physical effects, such as overlapping absorption from clumps at different velocities, differences in ionization state or optical depth, or weak emission components. 
Similar deviations from pure absorption have been reported in previous studies of Cyg~X-1's wind, including possible P~Cygni-type signatures at other orbital phases, 
e.g., \citet{Miskovicova2016-zi}. 
Although our observation includes the mid-dip phase ($\phi_{\rm orb} \sim 0.0$--0.1), even modest emission components outside dips could subtly skew the line profile. 
In this work, we focus on quantifying the basic parameters of the absorption lines -- such as equivalent width, centroid energy, and velocity shift -- using a simple model. 
More detailed modeling of the line profiles, accounting for these effects, will be pursued in future work.
With approximately 400 counts per 2~eV bin at 6.7~keV -- well above the threshold of 20 counts per bin recommended for $\chi^2$ statistics 
\citep{Yamada2019-rt} -- 
the fit was evaluated using the chi-squared method. This choice has no practical impact on the current analysis.

To investigate whether the redshift and turbulent velocity are significantly greater than zero, 
we performed spectral fits using four different models, as shown in Figure~\ref{fig_avespec_fit_fine}. 
The four cases considered were:  
(i) redshift $z = 0$ and turbulent velocity $v_{\rm turb} = 0$,  
(ii) $z = 0$ and $v_{\rm turb}$ free,  
(iii) $z$ free and $v_{\rm turb} = 0$,  
(iv) $z$ free and $v_{\rm turb}$ free.  
As shown in Figure~\ref{fig_avespec_fit_fine}, 
cases (i) and (ii) did not provide a good fit to the H-like K absorption lines, although case (ii) showed a slight improvement. 
In contrast, cases (iii) and (iv) improved the fits to a similar degree.
The detailed fitting parameters are listed in Table~\ref{table_avespec_fit}.  
According to the instrument team’s recommendation, the uncertainty in the energy scale 
is represented by 0.3~eV added in quadrature with the uncertainty from the in-orbit calibration pixel during the observation \citep{Megan2025}.
The obtained redshift, $-6 \times 10^{-4}$, corresponds to a blueshift of $-177$~km~s$^{-1}$, 
or $\sim$4~eV at 6.7~keV, which exceeds the estimated systematic uncertainty.  
This suggests that the highly ionized wind is outflowing toward the observer.
The proper motion of the system, assuming its association with Cyg~OB3, is estimated to be $11.4 \pm 3.0$~km~s$^{-1}$ \citep{Rao2020-lq}, 
which is lower than the energy-scale systematic uncertainty \citep{Megan2025}. 
The XRISM satellite had a maximum velocity of 7.59~km~s$^{-1}$ during the observation, with a maximum projected velocity of 6.72~km~s$^{-1}$ 
in the direction of Cyg~X-1, so its impact on the observed redshift is negligible.

An F-test was performed, yielding a statistic of 46.14 and a probability of $2.96 \times 10^{-20}$ when comparing case (i) and case (iv), 
indicating a statistically significant improvement. However, caution is warranted due to potential systematic effects associated with the single-zone model approximation. 
The constraint on the turbulent velocity remains uncertain due to degeneracies with redshift and the possible presence of multiple ionized absorbers, 
which complicate the interpretation.
We also note that the 0.5--3~keV count rate in the later part of the dipping phase ($\phi_{\rm orb} \sim 0.95$--0.17) 
is slightly higher than in the preceding non-dip phase ($\phi_{\rm orb} \sim 0.65$--0.85), suggesting the presence of mild intrinsic spectral evolution during the observation. 
This may be explained by a small increase in the overall soft X-ray emission from the source, possibly due to enhanced disk blackbody emission or a modest rise in accretion rate. Such intrinsic changes could affect the observed absorption features and spectral parameters. 
A time-resolved spectral analysis that simultaneously models both absorption and intrinsic continuum variability 
will be presented in a forthcoming paper.

\begin{figure}[h]
  \begin{center}
 \includegraphics[width=0.85\linewidth]{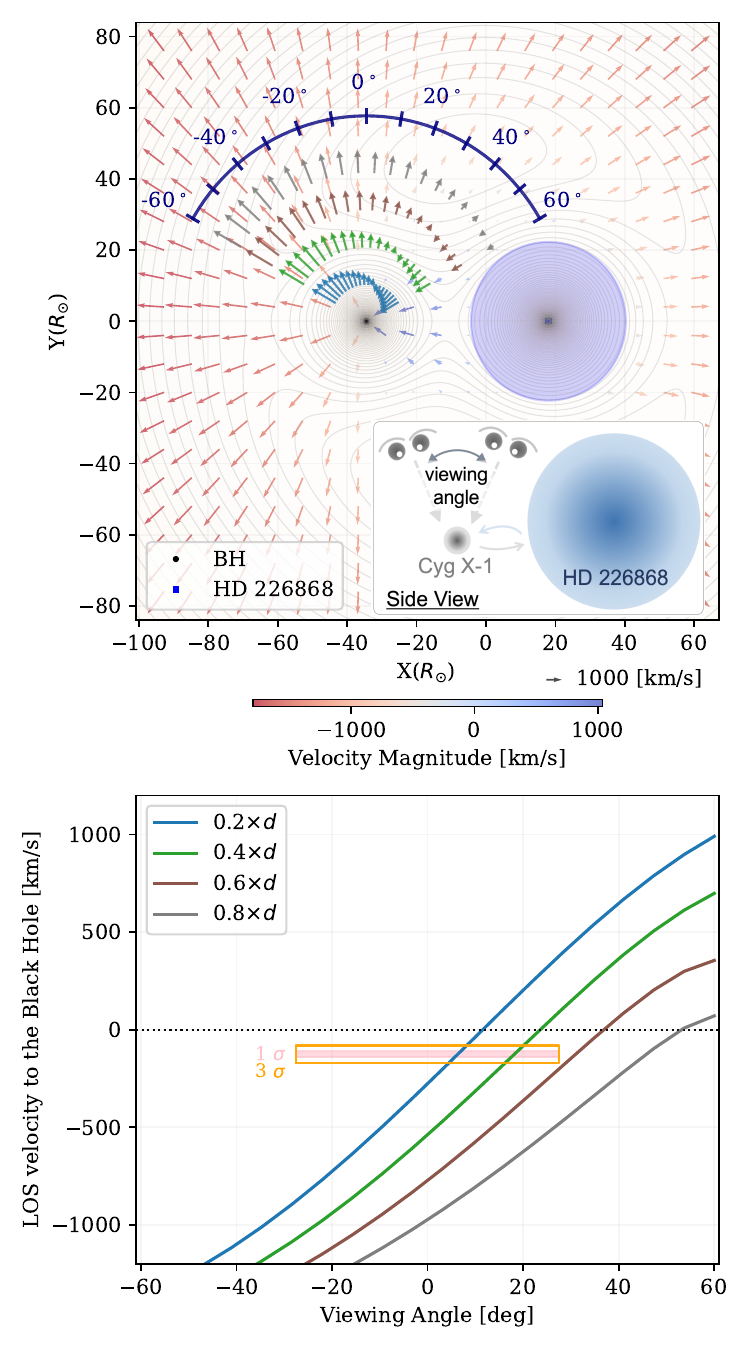} 
  \end{center}
  \caption{(Top) Visualization of the projected velocity along the LOS toward the BH, 
  assuming a spherically symmetric stellar wind.  
  The inset in the lower right illustrates a conceptual diagram, where the viewing angle is defined as 0 degrees 
  when the observer looks perpendicular to the line connecting the star and the BH.  
Velocity vectors are defined as positive in the direction toward the BH (shown as blue arrows), with arrow length indicating the magnitude.  
The assumed size of the companion star is depicted in blue, marked by a blue dot, while the BH is indicated in black.  
Arrows show the magnitude of the LOS velocity at various distances from the BH, with colors representing different radii: blue, green, brown, and gray correspond to 0.2$d$, 0.4$d$, 0.6$d$, and 0.8$d$, respectively.  
Roche potential contours are overlaid.
(Bottom)  
Dependence of the LOS velocity on distance from the BH, compared with spectral fitting results during dipping events.  
Colors indicating radial distance are consistent with those in the top panel.  
The 1$\sigma$ error range of the redshift measured during dips is shown in pink, and the 3$\sigma$ range in orange.  
The horizontal extent of the error boxes reflects the inclination range inferred from optical observations.
Alt text:  
(Top) Projected LOS velocity around a BH assuming a spherically symmetric wind.  
(Bottom) LOS velocity vs. distance from the BH with observed redshift constraints during dips.
}
\label{fig_ponti}
\end{figure}

\subsection{Spectra before and after $\phi_{\rm{orb}}=0.9$}

We divided the observation into ``dipping'' and ``non-dipping'' phases and explored the consequences 
on the spectral fit parameters and thus the implied physical conditions in the binary system.
The observation is divided into two: before and after $\phi_{\rm{orb}}=0.9$, and 
they are called `non-dip' and `dipping' phase.  
Figure \ref{fig_dip_nondip} (top) shows the spectra of Resolve during the non-dip and dipping phase. 
To visualize the absorption features, we fitted the powerlaw model with the non-dip spectrum, 
and took the ratios of the observed spectra to the model, and shown in Figure \ref{fig_dip_nondip} (middle),  
The ratio at 6.7 keV (He-like Fe K$_\alpha$) is nearly 0.5 with its width of $\sim$5 eV, 
which means the equivalent width (EW) of $\sim$ 2.5 eV. 
As a simple quantification of EW of the absorption lines,
we performed a fit in the 6.5--7.5 keV range for the dipping spectrum using a model of power-law with redshifted three negative Gaussian models. 
The three Gaussian components correspond to He-like Fe K$\alpha$, H-like Fe Ly$\alpha 2$, and Ly$\alpha 1$. 
As a result, the obtained values of EW were 2.3 eV, 0.4 eV, and 1.2 eV, respectively, 
which are broadly in line with the values expected from the spectral ratios.

We then performed spectral fits using a single ionized absorption model, \texttt{tbabs*cloudy\_abs*powerlaw}, over the 6.5--10~keV range for both the dip and non-dip spectra. 
The fitting results are summarized in Table~\ref{table_dip_nondip}. 
Due to the limited spectral features and lower statistics in the non-dip phase, 
some parameters have larger uncertainties compared to those from the dipping phase. 
While the column density in the non-dip phase cannot be well constrained under the single-zone approximation (with error bars including zero), 
there are hints of weak absorption and possible emission features. 
A detailed analysis of these structures will be presented in a future publication. 
The data, best-fit models, and residuals are shown in Figure~\ref{fig_dip_nondip}.

From the dip spectrum, we obtain a blueshift of approximately 100~km~s$^{-1}$ for the Fe~\textsc{xxvi} lines. 
This value is comparable to those reported by \citet{Miskovicova2016-zi} and \citet{Hanke2009-rh}, 
who found moderate velocity shifts (blueshifts or even slight redshifts) of a few $\times 10^1$--$10^2$~km~s$^{-1}$ at similar orbital phases ($\phi_{\rm orb} \sim 0.0$). 
Given XRISM's larger effective area in the Fe~K band compared to \textit{Chandra} HETG, 
our measurement of the Fe~\textsc{xxvi} line shift achieves improved statistical precision.
Our measured redshift is thus consistent with a nearly stationary or mildly outflowing wind along the line of sight near superior conjunction. 
In contrast, \citet{Miskovicova2016-zi} reported clear P~Cygni profiles at $\phi_{\rm orb} \sim 0.5$, 
indicating a denser wind with both absorption and redshifted re-emission. 
In our data, which correspond to superior conjunction, the line profiles are dominated by absorption, consistent with this phase-dependent behavior.
The derived ionization parameter ($\log \xi \sim 3$) and column density during the dip phase 
are also consistent with a nearly fully ionized wind, as reported in previous Chandra HETG studies. 

The observed turbulent velocity $v_{\text{turb}}$ can be compared to the expected thermal velocity, 
given by $v/c = \sqrt{2/A} \times \sqrt{kT / m_p c^2}$, where $A$ is the atomic mass number, $T$ is the plasma temperature, 
$k$ is the Boltzmann constant, and $m_p$ is the proton mass. 
For $A = 56$ and $T = 7 \times 10^7$~K, the thermal velocity is $v \approx 144$~km~s$^{-1}$, 
which is consistent with our fitted turbulent velocity, suggesting a plausible thermal origin (cf. Appendix~\ref{fig_xi}).
We also note a potential P~Cygni-like profile in the He-like Fe~K$\alpha$ line, although the statistical significance is currently limited. 
In addition, the dip spectrum exhibits an absorption feature consistent with Fe~\textsc{XXIV}~q at a rest energy of 6662.240~eV, 
indicating the presence of a lower ionization component. 
This will also be investigated in more detail in future work.

\begin{table}[h]
\caption{Fitting results of the single-zone absorber model for the Resolve spectra over 6.5 -- 10 keV during non-dip and dipping phases. 
The notes on the parameter are the same as used in table \ref{table_avespec_fit}.}
  \centering
  \begin{tabular}{ccc}
  \hline\hline
      paramater & dipping & non-dip  \\ \hline
      $\xi$ & 3.17$\pm$ 0.03 & 3.26$_{-0.11}^{+0.12}$  \\ 
      $N_{\rm H}$ [10$^{22}$cm$^{-2}$] & 0.73$\pm$ 0.07 & 0.15$_{-0.15}^{+0.04}$  \\[0.25em] 
      $z \times 10^{-4}$ & $-4.2\pm0.5$ & $-11.8_{-1.4}^{+1.5}$  \\ 
      $cz$ [km s$^{-1}$] & $-125.9\pm15.0$ & $-353.8_{-42.0}^{+45.0}$  \\[0.25em] 
      $v_{\rm{turb}}$ [km] & 141$_{-22}^{+23}$ & 0 (fix)$^a$  \\ 
      Photon Index & 1.95$\pm$ 0.02 & 2.05$\pm$ 0.02   \\ 
      norm & 1.81$\pm$ 0.07 & 2.29$_{-0.08}^{+0.09}$  \\ 
      \hline
      $\chi^2$/dof & 1895.2/1743 & 1780.6/1744 \\ 
  \hline\hline
  \end{tabular}
\label{table_dip_nondip}

\begin{small}
\begin{itemize}
\setlength{\parskip}{0cm} %
\setlength{\itemsep}{0cm} %
\item[$^a$] $v_{\rm{turb}}$ was fixed at 0 due to the lack of distinct absorption features in the data.
\end{itemize}
\end{small}

\end{table}

\section{Discussion}
We obtained XRISM observations of Cyg~X-1 in the low/hard state,  
revealing detailed profiles of He-like and H-like Fe-K$\alpha$ absorption lines.  
We analyzed their properties in two orbital phase intervals, spanning from the non-dip phase ($\phi_{\rm{orb}} \sim 0.66$--0.85) 
to the dipping phase ($\phi_{\rm{orb}} \sim 0.90$--0.15).  
Although these absorption lines were previously detected with \textit{Chandra} HETG,  
the superior spectral resolution and effective area of XRISM/Resolve allow for more definitive detections in the Fe~K band.  
While a time-resolved analysis of line evolution throughout the dip will be presented in a forthcoming publication,  
our results demonstrate the capability of XRISM to capture subtle changes in absorption properties across different orbital phases.  

To study the spectral evolution, we divided the observation into two phases and quantified the changes in their spectral properties. 
Using conventional spectral models, we found that the column density of the He-like Fe-K$\alpha$ absorption line increased 
by a factor of $\sim 5$ over a timescale of a few tens of ks.  
This variation likely reflects changes in the ionized absorber as the system moves from the non-dip to the dipping phase, 
consistent with studies of variability on much shorter time scales \citep{Lai2024-fu}. 
Such time variability is difficult to explain with a static absorbing medium and instead suggests dynamical changes in the stellar wind.  
Although a detailed analysis of short-term variability is beyond the scope of this work, 
the high time resolution and soft X-ray sensitivity of Xtend will enable future studies during dips.  
While it is evident from the performance of the instrument, 
this H-like Fe absorption with its EW of an $\sim$ eV are 
more clearly detected than that of $\sim$ 5 eV obtained with the Suzaku CCD spectra during its high/soft state \citep{Yamada2013-wa}. 

Our results are consistent with previous measurements. 
\citet{Hanke2009-rh} analyzed \textit{Chandra}/HETG spectra of Cyg~X-1 during a 16.1~ks non-dip interval at a similar orbital phase ($\phi_{\rm orb} \sim 0.93$--$0.03$) in the low/hard state, 
and reported a blueshift velocity of $116^{+501}_{-531}$~km~s$^{-1}$ from the He-like Fe~\textsc{xxv}~K$\alpha$ absorption line.
\citet{Hirsch2019-em} investigated three \textit{Chandra}/HETG observations of Cyg~X-1 during dips at similar orbital phases ($\phi_{\rm orb} \sim 0$) in the low/hard state, 
focusing on absorption lines from silicon and sulfur in the 1.6--2.7~keV range. 
They examined the distribution of Doppler shifts across multiple lines and found that the velocity distributions varied between observations: 
for ObsID~3814, the velocities clustered around 0~km~s$^{-1}$; for ObsID~8525, the velocities were more blueshifted, scattering around $-60$~km~s$^{-1}$; 
and for ObsID~9847, the distribution was centered near $-230$~km~s$^{-1}$, indicating a stronger blueshift. 
Our XRISM observation yields a blueshift of $125.9 \pm 15$~km~s$^{-1}$ for the Fe~\textsc{xxvi} line. 
While not a strictly direct comparison due to differences in ion species and dip phases, these values are in general agreement within uncertainties. 
We also obtain an ionization parameter of $\log \xi = 3.0 \pm 0.1$ and a column density of $\sim 3 \times 10^{21}$~cm$^{-2}$, 
which are consistent with those inferred from detections of H-like and He-like Fe absorption lines in \textit{Chandra}/HETG dip spectra.

Although detections of highly ionized Fe-K absorption lines remain relatively rare,
this may reflect differences in the absorbing medium -- such as more complete ionization driven by the increased number of ionizing photons in the high/soft state 
-- or changes in the global wind structure under different accretion conditions.
Recent studies have suggested that the overall density, geometry, and acceleration profile of the stellar wind in Cyg~X-1 vary between spectral states.
For example, \citet{Ramachandran2025-jn} suggest that the wind's terminal velocity and mass-loss rate differ between the hard and soft states,
indicating changes in the global wind structure depending on the ionizing flux.
Similarly, \citet{Grinberg2015-kl} showed that simple CAK wind models could not reproduce the observed orbital variability unless asymmetric structures such as focused winds or clumping were included.
Moreover, radiation-hydrodynamic simulations by \citet{Cechura2015-cc} explicitly demonstrate how the wind structure responds to changes in X-ray luminosity.
In the low/hard state, the hemisphere of the donor facing the compact object experiences reduced wind acceleration due to X-ray photoionization,
but still supports a dense, focused wind stream toward the compact object. In contrast, in the high/soft state, the intense X-ray radiation produces a fully ionized bubble 
that suppresses wind launching from the illuminated hemisphere. This disrupts the focused wind stream, alters the geometry of the bow shock, 
and leads to a substantial decrease in mass transfer. These simulations support the view that highly ionized absorption lines may preferentially appear in certain configurations.
In the low/hard state, the abundance of hard X-ray photons extending up to $\sim$100~keV may also contribute to the formation of extended, low-density, and highly ionized absorbing structures.
Such structures may arise from components such as an accretion stream, variations in the scale height of the X-ray emitting corona \citep{Wen1999-iw}, 
or precession of the accretion disk accompanied by an accretion bulge \citep{Poutanen2008-vn}.

The blue-shifted absorber detected in this study is examined under the assumption of the simplest model, a spherically symmetric stellar wind model. 
Considering the radial direction $r$ in a polar coordinate system centered on the star, 
the velocity of the stellar wind in the radial direction, $v(r)$, can be expressed as:
\begin{equation}
v(r) = v_{\infty} \cdot f\left(\frac{r}{R_{\star}}\right),
\end{equation}
where $v_{\infty}$ is the terminal wind velocity, and $f(x)$ represents the ratio of the velocity to the terminal velocity, given by \cite{Lamers1993-ev}:
\begin{equation}
f(x) = f_0 + (1 - f_0) \cdot (1 - 1/x)^\beta \quad (\text{for } x > 1). 
\end{equation}
In this model, the key parameters are $f_0$, which denotes the ratio of the base wind velocity $v_0$ to the terminal velocity $v_{\infty}$ $\left(f_0 = v_0 / v_{\infty}\right)$, and $\beta$, which determines the acceleration of the stellar wind. Using this model, we computed the velocity field by assuming $f_0 = 0.01$, $\beta = 0.75$, and $v_{\infty} = 2100$ km s$^{-1}$ \citep{Herrero1995-oh}. 

Figure \ref{fig_ponti} presents the velocity field projected along the observer's line of sight to the BH. 
The stellar radius was assumed to be $22.3 R_{\odot}$. Based on the orbital period and masses, 
the binary separation $d$ was set to $3.65 \times 10^{12}$ cm ($52.5 R_{\odot}$). 
As discussed in \citep{Hanke2009-rh}, the velocity component of the stellar wind in the direction of the BH 
becomes zero along the circle defined by the BH and the star as its diameter. 
Inside this circle, the flow transitions into an inflow toward the BH, as indicated by the blue arrows in figure \ref{fig_ponti} (top).
The effect of changing the distance from the BH from $0.2d$ to $0.8d$ on the velocity field was examined for inclinations 
within $\pm 60^\circ$, indicated by arrows. 
The lower part of the figure illustrates the relationship between the viewing angle and the LOS velocity toward the BH.

The observed blueshift ranges by fitting the dipping spectrum within $1\sigma$ (-141 to -111 km s$^{-1}$) and $3\sigma$ (-171 to -81 km s$^{-1}$)
are shown in Figure \ref{fig_ponti} (bottom), 
along with the inclination constraints derived from optical observations. 
If the stellar wind is truly spherically symmetric, a blueshift component of $\sim$100 km/s at $\sim 0.5 d$ would be detected 
if observed from an inclination of approximately $+30^\circ$ at an orbital phase of $\sim 0.9$--$0.1$. 
Assuming $L \sim 3 \times 10^{37}  \text{ergs}~\text{s}^{-1}$, $\xi \sim 10^{3.2}$, $N_{\rm H} \sim 7 \times 10^{21}$cm$^{-2}$, 
and the distance from the BH $\sim 0.5 d$, the density and its length are estimated to be $6 \times 10^{9} \text{cm}^{-3}$ and $\sim 10^{12} \text{cm}$, respectively.  
The X-ray intensity variations below 3 keV in figure \ref{fig:lc} occur on timescales 
as short as about a $\sim$ ks and have been reported to be as rapid as tens of seconds \citep{Lai2024-fu}, 
possibly reflecting density inhomogeneities in the absorbing material. 
This suggests that the high-density, low-ionization region near the BH exhibits more rapid variations, 
while the lower-density, highly ionized, more distant region varies more slowly. 
However, as extensively discussed in previous studies, the presence of focused and clumpy winds, or effect of partial covering (e.g., \cite{Hirsch2019-em}) may introduce additional complexities. 
More comprehensive observations of dipping phenomena across different spectral states will be crucial for constructing a complete picture of the stellar wind and accretion flow around the BH.

\section{Summary}

We conducted the first high-resolution X-ray spectral analysis of Cyg X-1 using XRISM,  
taking advantage of the superior energy resolution of the Resolve instrument to investigate highly ionized iron absorption lines.  
Our study provides detailed constraints on the ionization states, column densities, and velocities of the absorbing material,  
revealing significant temporal variations across different orbital phases.

The lightcurves from Xtend clearly show rapid variability during the dip.  
By dividing the observation into two phases, we tracked the spectral evolution and 
found a substantial increase in the column density of the He-like Fe-K$\alpha$ absorption line on a timescale of tens of ks.  
Through spectral modeling, we derived ionization parameters of $\xi \sim 3$, column densities on the order of $\sim 10^{21}$ cm$^{-2}$,  
and blueshifted absorption corresponding to outflow velocities of $\sim 100$ km~s$^{-1}$.  
These results suggest the presence of a dynamic and possibly inhomogeneous stellar wind that undergoes rapid variations  
in density and ionization conditions as it interacts with the compact object.

Although highly ionized Fe-K absorption lines have rarely been observed in the low/hard state,  
they appear to be associated with transient dipping events, which occur more frequently in this state than in the high/soft state.  
This difference may arise from variations in the density or ionization structure of the absorber.  
However, additional geometrical effects -- such as changes in the scale height of the corona, interactions with an accretion stream, or disk precession 
-- may also play a role in shaping the observed absorption profiles.

To interpret the blueshifted absorption, we adopted a simple spherically symmetric stellar wind model  
and examined the projected velocity field along the observer's LOS.  
Our calculations indicate that a blueshift of $\sim 100$ km~s$^{-1}$ is expected for an inclination of $\sim 30^\circ$  
at an orbital phase of $\sim 0.9$--$0.1$, assuming a smooth stellar wind.  
However, as discussed extensively in previous studies, deviations from spherical symmetry -- such as focused or clumpy winds -- 
likely contribute to the observed variability.
Our results highlight the importance of continued high-resolution X-ray observations to systematically probe the stellar wind and accretion dynamics in Cyg X-1.  
Future monitoring campaigns using X-ray calorimeters will enable a more comprehensive understanding of the wind-fed accretion process,  
allowing us to refine models of BH accretion in high-mass X-ray binaries.

\begin{ack}
This observation was finally realized through the longstanding contributions of the XRISM team members, engineers in detectors and space engineering, and the operation team. The efforts toward the Hitomi observation of Cyg X-1, planned for March 30, 2016, are respectfully acknowledged.
The authors acknowledge the memory of the late Dr. Jeffrey E. McClintock, Dr. Magnus Axelsson, and Dr. Katja Pottschmidt, whose encouragement and support remain deeply valued.
\end{ack}

\section*{Funding}
Part of this work was supported by JSPS KAKENHI grant Nos.\ JP24K00672 (S.Y.), 	23K22543 (S.Y.), JP21K13958 (M.M.), 
23K25882 and 23H04895 (T.M.), 22H01269 (T.K.), Yamada Science Foundation (M.M.), and 
JSPS Core-to-Core Program (grant number:JPJSCCA20220002).

\section*{Data availability} 
The XRISM data underlying this article are available in heasarc data repository (NASA/GSFC) or DARTS (JAXA/ISAS). All the data reduction tools are available as a ftools package. 

\newpage

\appendix 

\section{Photoionization with \texttt{Cloudy}}
\label{sec_cloudy}

To assess the ionization state of a low-density plasma irradiated by a compact object, we performed photoionization calculations using the spectral synthesis code \texttt{Cloudy}. 
As the incident spectrum, we adopted a SED representative of the low/hard state of BH binaries, 
composed of a multicolor disk blackbody and thermal Comptonization as described in \ref{sec:timeave}. 
The hydrogen density was fixed at $n_{\mathrm{H}} = 10^{10}$ [cm$^{-3}$], 
The ionization parameter $\xi$ was varied in the range $\log \xi = -1$ to $4$, 
with the calculation limited to a single zone to focus on the equilibrium properties near the illuminated surface.

\begin{figure}[h]
  \begin{center}
    \includegraphics[width=0.9\linewidth]{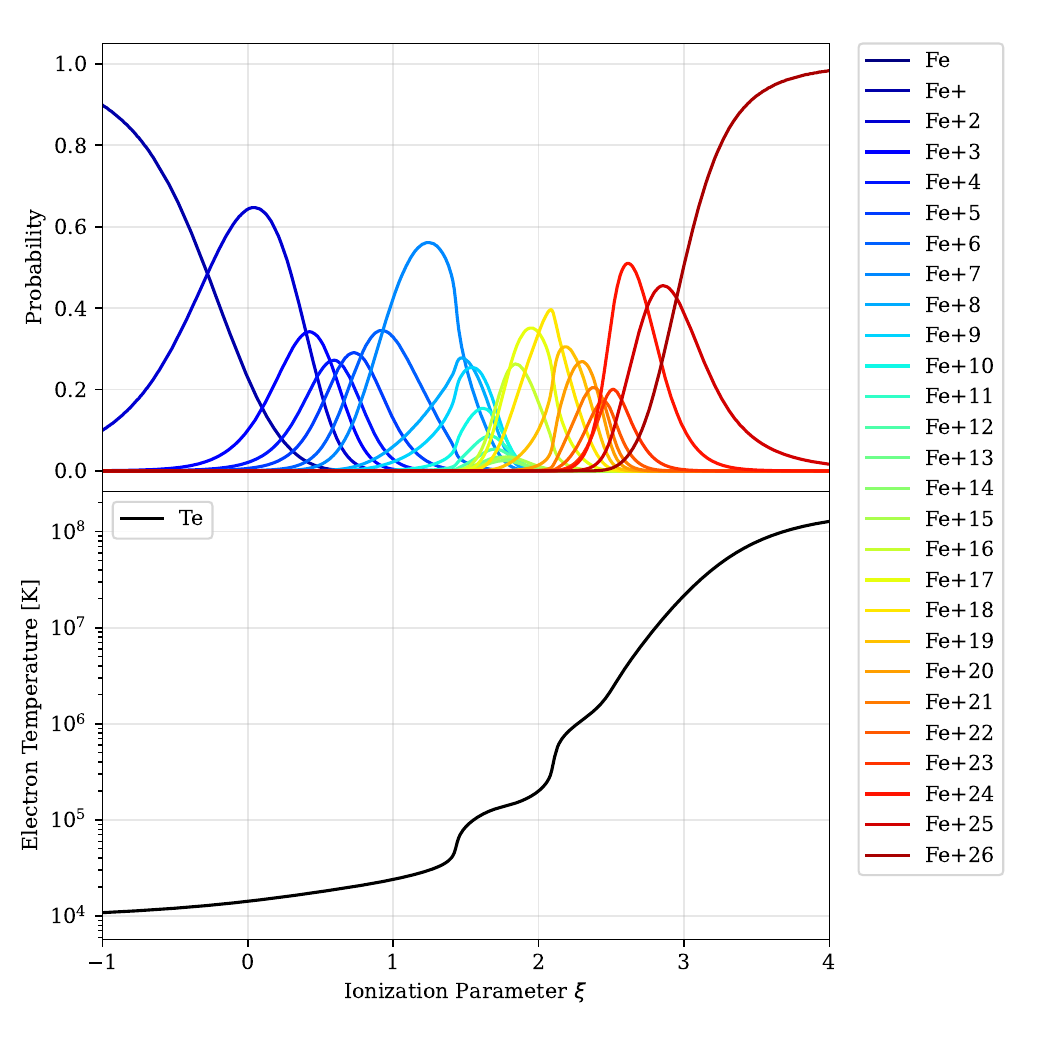} 
  \end{center}
  \caption{Results of the photoionization calculations using \texttt{Cloudy} for the BH low/hard state SED. 
  The horizontal axis represents the ionization parameter $\log \xi$.
Top panel: Fractional abundances of iron ions (Fe, Fe$^+$, ..., Fe$^{25+}$) as a function of $\log \xi$.
Bottom panel: Electron temperature as a function of $\log \xi$.
At $\log \xi \gtrsim 3$, iron is nearly fully ionized, and the electron temperature exceeds $10^7$ K.
Alt text:  $\log \xi$ v.s. probability of the ion species and electron temperature. 
}
\label{fig_xi}
\end{figure}

Figure~\ref{fig_xi} shows the distributions of fractional ion abundances of iron (upper panel) and the electron temperature (lower panel) 
as a function of the ionization parameter $\xi$. 
For $\log \xi \lesssim 0$, iron remains mostly neutral, whereas for $\log \xi \gtrsim 3$, even the heaviest iron ions become nearly fully ionized. 
In this high-ionization regime, the electron temperature exceeds $10^7$~K, as illustrated in figure~\ref{fig_xi}. 
Given that the ionization parameter inferred from our spectral fits is slightly above $\log \xi = 3$, the absorption features observed in this observation are likely produced by a hot, highly ionized plasma.


\section{Energy Calibration Using Fe-55 Isotopes}
\label{sec_fe55}

\begin{figure}[h]
  \begin{center}
    \includegraphics[width=0.9\linewidth]{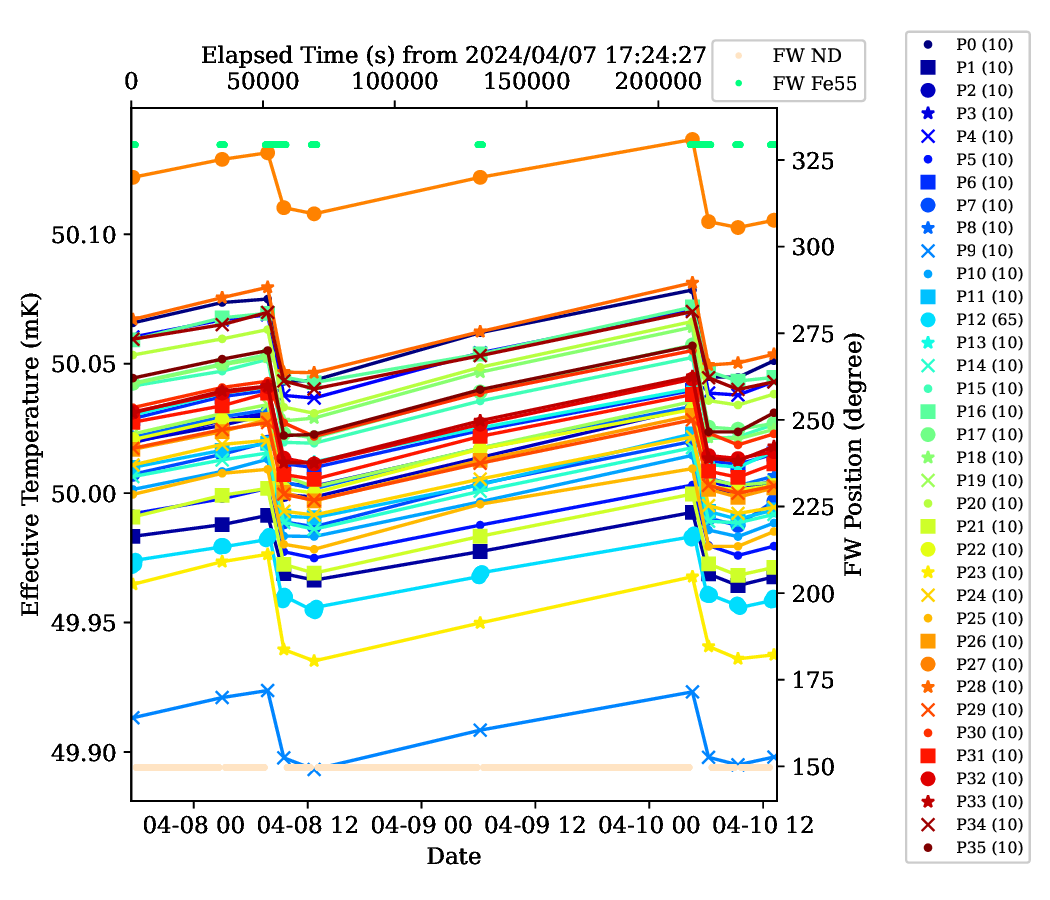} 
  \end{center}
  \caption{Relationship between the energy calibration points from Fe-55 source irradiation and the rotation angle of the filter wheel. Effective temperature anchor points used for energy calibration across all pixels are displayed with different colors and markers for each pixel. The numbers in parentheses in the legend indicate the number of calibration points for each pixel. For non-calibration pixels, all have 10 calibration points. The right axis of the figure shows the rotation angle of the filter wheel, with 150 degrees corresponding to the position of the neutral density filter (beige) and 330 degrees corresponding to the Fe-55 source position (yellow-green).
Alt text: Fe-55 source calibration points v.s. filter wheel rotation angle. }
\label{fig_ghf}
\end{figure}

During this observation, XRISM/Resolve energy calibration was performed using ten irradiations of the onboard $^{55}$Fe source. 
Two ADR recycle events caused notable energy scale shifts. Because the $^{55}$Fe calibration was conducted during target observation, 
the number of usable Hp events in the central four pixels was limited. 
To compensate, we set \texttt{rslgain} parameters to \texttt{min = 60} and \texttt{dshift = 12} to generate an appropriate gain history file. 
Event files were then reprocessed with \texttt{rslpha2pi}. 
The resulting gain history and effective temperatures are shown in Fig.~\ref{fig_ghf}; see \citet{Eckart2025-al} and \citet{Porter2025-mo} for method details. 
The gain variation introduced by reprocessing was within the statistical uncertainties and acceptable for standard analysis.

\begin{figure}[h]
  \begin{center}
    \includegraphics[width=0.9\linewidth]{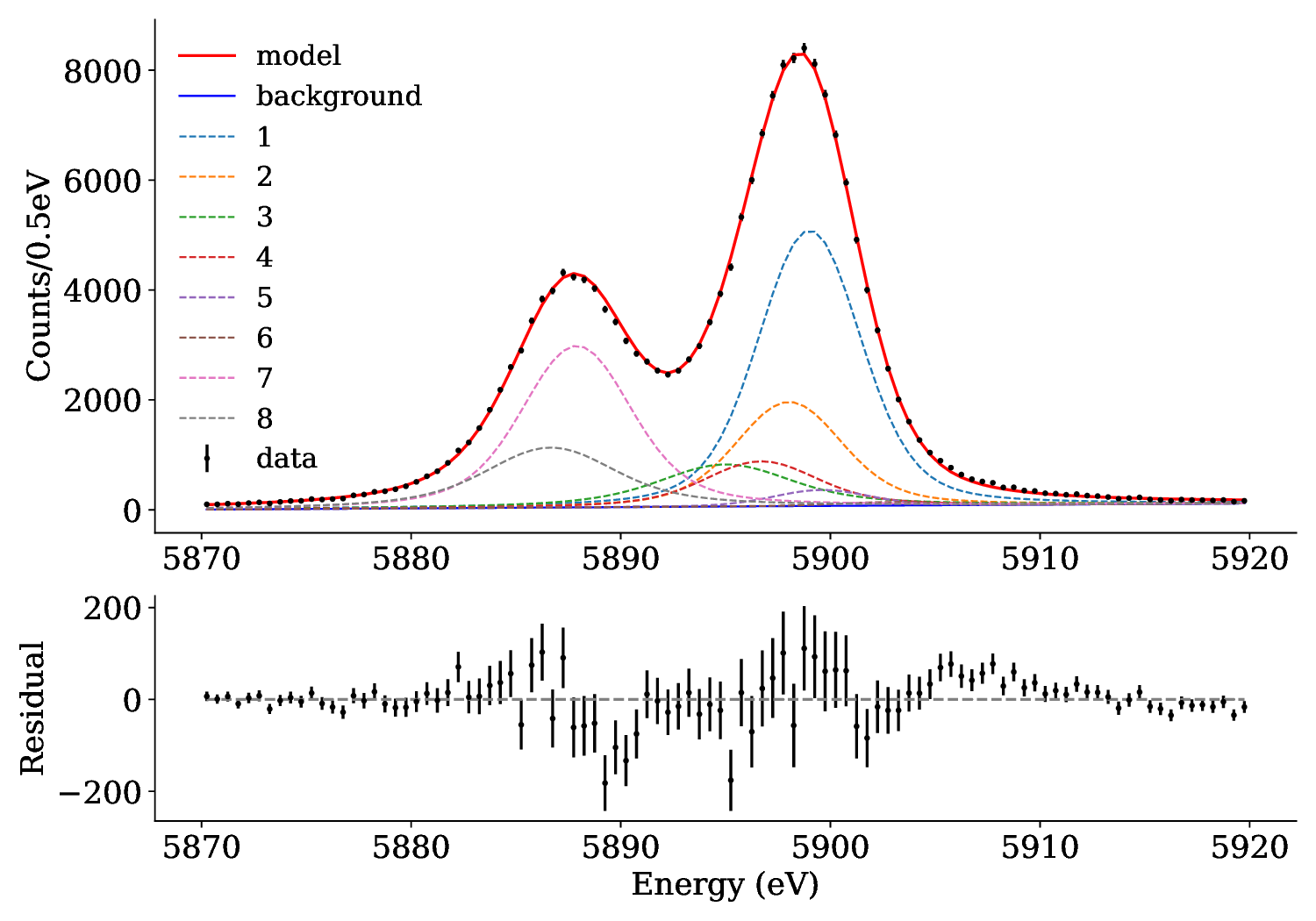} 
  \end{center}
  \caption{
Spectrum of Fe-55 isotopes obtained during the observation of Cyg X-1, 
summed over all pixels except the calibration pixel, is shown together with the best-fit model. 
For the theoretical modeling of Mn K$\alpha$, the total model is shown in red, while the eight constituent Voigt profiles are 
displayed in different colors. 
The lower panel shows the residuals between the data and the model.  
Alt text: Energy vs. count rate spectrum of Fe-55 isotopes with the best-fit model overlay.
}
\label{fig_mnka}
\end{figure}

\begin{figure}[h]
  \begin{center}
    \includegraphics[width=0.9\linewidth]{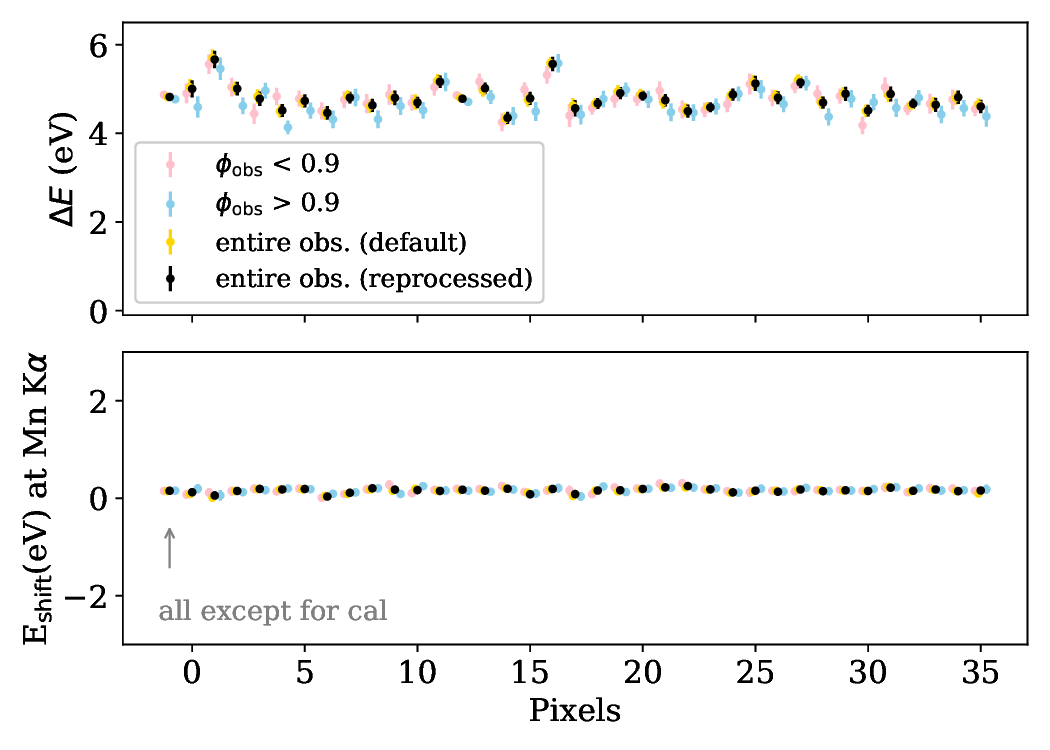} 
  \end{center}
  \caption{
Evaluation of energy resolution (top) and energy centroid (bottom) using the Mn K$\alpha$ line from Fe-55 isotopes.  
Results are shown in pink and light blue for $\phi_{\rm{orb}} < 0.9$ and $\phi_{\rm{orb}} \geq 0.9$, respectively;  
yellow indicates the entire observation, and black shows the result from data reprocessed with \texttt{rslgain}.  
In both panels, the position at pixel $-1$ indicates the fit result obtained by summing over all pixels except the calibration pixel.  
Alt text: Energy resolution (top) and energy centroid (bottom) vs. pixel number.}
\label{fig_mnka_result}
\end{figure}

Figure~\ref{fig_mnka} shows the $^{55}$Mn K$\alpha$ complex spectrum (from $^{55}$Fe decay) summed over all pixels except the calibration pixel. 
This line complex, centered near 5.9~keV, consists of multiple Voigt profiles derived from high-resolution ground-based measurements \citep{Holzer1997-cr}, 
with parameters from \citet{Yamada2019-rt}.
We evaluated energy resolution and energy scale through spectral fitting, as summarized in Fig.\ref{fig_mnka_result}. 
Four datasets were analyzed: the full observation, the reprocessed data, and subsets divided at orbital phase $\phi_{\rm orb} = 0.9$. 
For each, per-pixel fits are shown, with summed results (excluding the calibration pixel) labeled as pixel $-1$. 
We found an energy resolution of $\sim$4.7eV and scale offsets below 0.3~eV, confirming that the measured blueshift of 100~kms$^{-1}$ ($\sim$2eV at 6.7~keV) exceeds the calibration uncertainty (see Section 5 in \citet{Eckart2025-al}).

\section{Detailed on Grade Branching Ratios}
\label{sec:grade}
\begin{figure}[h]
  \begin{center}
    \includegraphics[width=0.9\linewidth]{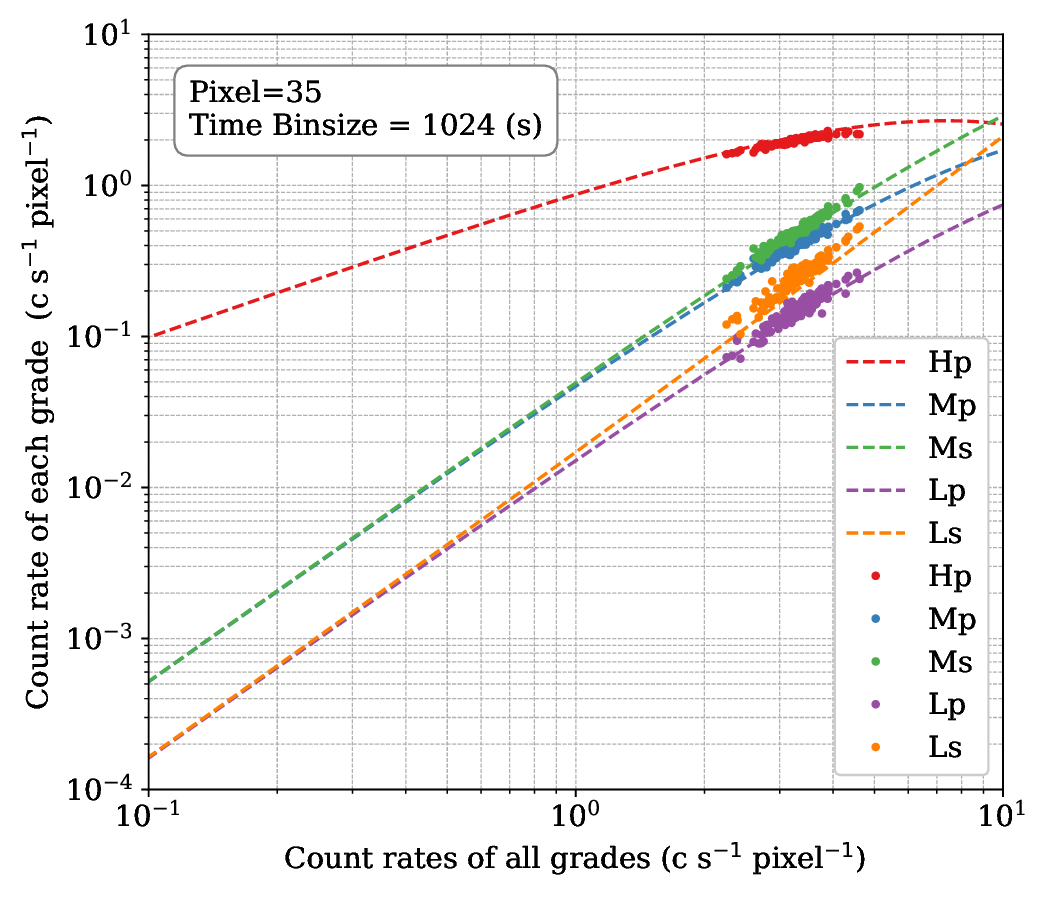} 
  \end{center}
  \caption{Relationship between count rates and grade branching ratios. The horizontal axis represents the total count rate for all grades in pixel 35, while the vertical axis shows the count rate for each grade. The count rates were calculated over intervals of 1024 seconds. The theoretical curves are shown as dotted lines.
Alt text: Total count rate v.s. grade branching ratios in pixel 35.}
\label{fig_bratio}
\end{figure}

For bright sources, it is essential to account for the grade branching ratios in the analysis, 
and to ensure that the observed grade distribution is consistent with the expected model.
For instance, when the count rate becomes excessively high, the proportion of Hp events may decrease, while secondary events increase. This phenomenon can result in a decrease in the Hp count rate, even though the source is intrinsically brighter. To illustrate the conditions during this observation, we present the total event rate and branching ratios for pixel 35, one of the central four pixels, in figure \ref{fig_bratio}.

The total event rate for pixel 35 is on the order of a few Hz, where Hp events are most abundant, and the Hp count rate increases slightly with brightness. For pixels outside the central four, the total event rates are lower, Hp events still dominate, and the linearity of the Hp count rate is better preserved. This observation was therefore conducted within a regime where the linearity between input event rates and detected rates was well maintained. It should be noted, however, that an ND filter was used in this observation, which must be taken into account when comparing with data obtained without an ND filter or when the gate valve is opened; in such cases, high-count-rate effects should be carefully considered, as discussed by \citet{Mizumoto2025-xb}.

\bibliographystyle{aasjournal}
\bibliography{Cyg-X-1_cvt,XRISM_cvt,ASTRO-H_cvt,Calorimeter_cvt,Calorimeter_cvt_spie,Plasma-Physics_cvt}
\end{document}